\documentclass[reprint,aps,superscriptaddress,amsmath,amssymb,longbibliography]{revtex4-1}
                                                                                                                                                                                                                                                                                                                                                                                                                                                                                                                                                                                                                                                                                                                                                                                                                                                                                                                                                                                                                                                                                                                                                                                                                                                                 
\usepackage{graphicx}
\usepackage{dcolumn}
\usepackage{bm}
\usepackage{color}
\usepackage{xspace}
\usepackage{ulem}
\usepackage{xfrac}
\usepackage[colorlinks=true,urlcolor=blue,citecolor=blue,linkcolor=blue,bookmarks=false,pdfstartview={FitH}]{hyperref}

\def \cm-1{cm$^{-1}$\,}

\begin{document} 
\title{Charge density wave order in the kagome metal AV$_3$Sb$_5$ (A= Cs, Rb, K)}
                                
\author{Shangfei~Wu}
\email{sw666@physics.rutgers.edu}
\affiliation{Department of Physics and Astronomy, Rutgers University, Piscataway, New Jersey 08854, USA}
\author{Brenden~R.~Ortiz}
\affiliation{Materials Department and California Nanosystems Institute, University of California Santa Barbara, Santa Barbara, California 93106, USA}
\author{Hengxin~Tan}
\affiliation{Department of Condensed Matter Physics, Weizmann Institute of Science, Rehovot 7610001, Israel}
\author{Stephen~D.~Wilson}
\affiliation{Materials Department and California Nanosystems Institute, University of California Santa Barbara, Santa Barbara, California 93106, USA}
\author{Binghai~Yan} 
\affiliation{Department of Condensed Matter Physics, Weizmann Institute of Science, Rehovot 7610001, Israel}
\author{Turan~Birol} 
\affiliation{Department of Chemical Engineering and Materials Science, University of Minnesota, MN 55455, USA}
\author{Girsh~Blumberg} 
\email{girsh@physics.rutgers.edu}
\affiliation{Department of Physics and Astronomy, Rutgers University,
Piscataway, New Jersey 08854, USA}
\affiliation{National Institute of Chemical Physics and Biophysics,
12618 Tallinn, Estonia}
\date{\today}               
                                                                                                                                                                                                                                                   
\begin{abstract}  
                                                          
We employ polarization-resolved electronic Raman spectroscopy and density functional theory to study the primary and secondary order parameters, as well as their interplay, in the charge density wave (CDW) state of the kagome metal AV$_3$Sb$_5$.  
Previous x-ray diffraction data at 15\,K established that the CDW order in CsV$_3$Sb$_5$ comprises of a $2\times2\times4$ structure: one layer of inverse-star-of-David and three consecutive layers of star-of-David pattern.    
We analyze the lattice distortions based on the $2\times2\times4$ structure at 15\,K, and find that the $U_1$ lattice distortion is the primary-like (leading) order parameter while $M^+_1$ and $L^-_2$ distortions are secondary-like order parameters for vanadium displacements.  
This conclusion is confirmed by the calculation of bare susceptibility $\chi'_0(q)$ that shows a broad peak at around $q_z=0.25$ along the hexagonal Brillouin zone face central line ($U$ line).
We also identify several phonon modes emerging in the CDW state, which are lattice vibration modes related to V and Sb atoms as well as alkali-metal atoms.    
The detailed temperature evolution of these modes' frequencies, half-width at half-maximums, and integrated intensities support a phase diagram with two successive structural phase transitions in CsV$_3$Sb$_5$: the first one with a primary-like order parameter appearing at $T_S=94$\,K and the second isostructural one  appearing at around $T^*=70$\,K. Furthermore, the $T$-dependence of the integrated intensity for these modes shows two types of behavior below $T_S$: the low-energy modes show a plateau-like behavior below $T^*$ while the high-energy modes monotonically increase below $T_S$. These two behaviors are captured by the Landau free-energy model incorporating the interplay between the primary-like and the secondary-like order parameters via trilinear coupling. Especially, the sign of the trilinear term that couples order parameters with different wave-vectors determines whether the primary-like and secondary-like order parameters cooperate or compete with each other, thus determining the shape of the $T$ dependence of the intensities of Bragg peak in x-ray data and the amplitude modes in Raman data below $T_S$.
These results provide an accurate basis for studying the interplay between multiple CDW order parameters in kagome metal systems. 
                                                                                                                                    
\end{abstract}
                                                                                
                                                                                                                                                                                        
\maketitle
           
\section{Introduction}    
                                                                     
The kagome lattice is a model system to study the electronic and magnetic properties~\cite{Itiro1951,Broholmeaay0668}. 
The corner shared triangle network of the kagome lattice enables three sublattice interferences, which give rise to a variety of exotic physics, for example, flat bands, van Hove singularities, Dirac dispersions in its electronic structure, frustrated magnetism, to name a few. 
Various electronic orders such as charge and spin density wave order, charge bond order, chiral flux order, nematic order, and superconductivity are under rigorous investigations~\cite{Wang2013PhysRevB,Kiesel2013PhysRevLett,FENG_2021SB,Denner_arxiv2021,Lin_arxiv2021,Park_arxiv2021,Setty_arxiv2021,Feng2_arxiv2021,Christensen_arxiv2021,Tazai_arxiv2021,Jiang_arxiv2021_review,Xie_neutron_arxiv2021,Chen_arxiv2021,Liang_arxiv2021}.                                                                              
                                                                                                                                                             
\begin{figure*}[!ht] 
\begin{center}
\includegraphics[width=2\columnwidth]{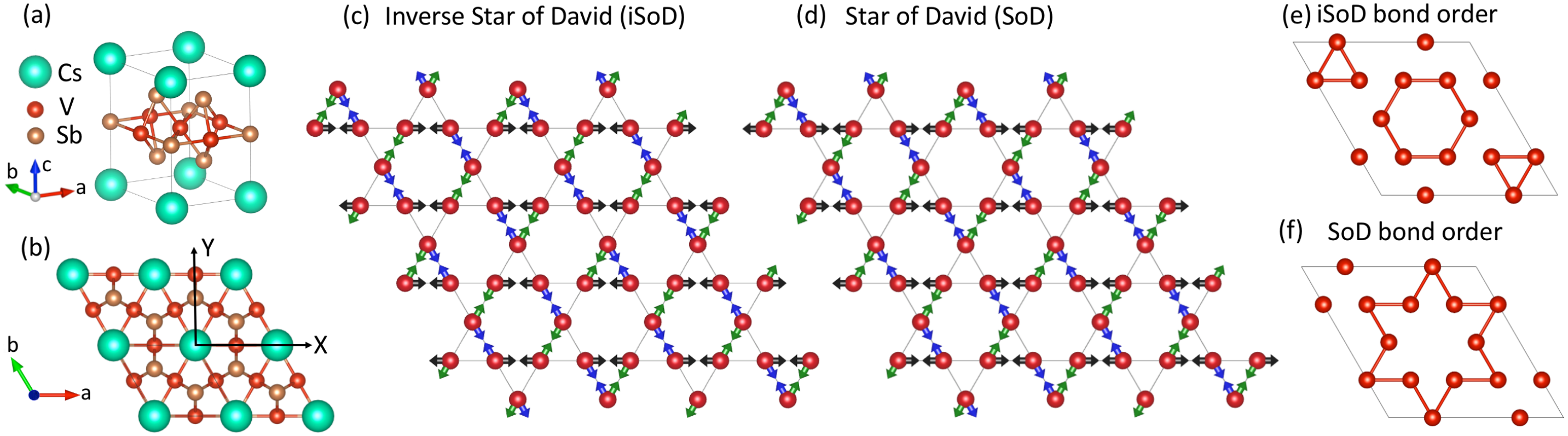}
\end{center}
\caption{\label{Fig1_structure} 
(a) Crystal structure of CsV$_3$Sb$_5$ in the high-temperature phase (space group: $P6/mmm$, No.~191, point group: $D_{6h}$). 
(b) Top view of the high-temperature crystal structure along the $c$ axis and definitions of $X$ and $Y$ directions.  
(c) The inverse Star of David (iSoD) $2\times2\times1$ CDW phase. The black, blue, and green arrows represent the three V-V displacement directions. (d) The Star of David (SoD) $2\times2\times1$ CDW phase, which is obtained by merely reversing the direction of black arrows in (c). (e) The shorter V-V bond pattern for iSoD structure. (f) Same as (f) but for SoD structure.
}
\end{figure*}

Recently, a three-dimensional charge density wave (CDW) order, which coexists with superconductivity (SC) at low temperatures, was discovered in AV$_3$Sb$_5$ (A= Cs, Rb, K) kagome metals~[Fig.~\ref{Fig1_structure}(a)]~\cite{FENG_2021SB,Ortiz2019PhysRevMaterials,Ortiz2020PhysRevLett,Ortiz2021PhysRevMaterials,Jiang2021NatureMaterial,Tan_arxiv2021,Zhao_arxiv2021,Chen_arxiv2021,Li_arxiv2021,ChenPhysRevLett.126.247001}. 
Superconductivity emerges at $T_c=$1$\sim$3\,K, much lower than the CDW transition temperature ($T_S=$80$\sim$100\,K)~\cite{Ortiz2019PhysRevMaterials,Ortiz2020PhysRevLett,Ortiz2021PhysRevMaterials}. Superconductivity competes with the CDW order in AV$_3$Sb$_5$, as $T_c$ increases when the CDW order is suppressed by hydrostatic pressure~\cite{Yu2021,ChenPhysRevLett.126.247001}, or by the hole doping~\cite{Oey_2021arXiv,Liu_Ti_2021arXiv,Song2021PhysRevLett}, or by thickness reduction~\cite{Song2021PhysRevLett,Song_2021arXiv}.
Furthermore, the CDW state shows a large extrinsic anomalous Hall effect~\cite{Yang2020_ScienceAdvances,Yu2021PhysRevB} in the absence of magnetic ordering~\cite{Kenney_2021JPCM}. In the same CDW  state, muon spin relaxation studies revealed a striking enhancement of the internal field just below $T_S$ which persists into the SC state, suggesting time-reversal symmetry breaking~\cite{III_arxiv2021_usR,Yu_arxiv2021usR}.
Thus, clarification of the symmetry, the nature, and the low-temperature properties of the CDW order is pivotal for understanding the superconductivity in AV$_3$Sb$_5$ system.

The origin of the CDW order in AV$_3$Sb$_5$ remains under debate~\cite{Jiang2021NatureMaterial,FENG_2021SB,Denner_arxiv2021,Lin_arxiv2021,Park_arxiv2021,Setty_arxiv2021,Feng2_arxiv2021,Christensen_arxiv2021,Tan_arxiv2021,ZhouXJ_arxiv2021,Jiang_arxiv2021_review}.
However, there is a consensus that CDW order leads to lattice distortions, which mostly consist of the 
displacements of vanadium atoms in the kagome plane: the inverse-star-of-David (iSoD) type CDW [Fig.~\ref{Fig1_structure}(c)] and the star-of-David (SoD) type CDW [Fig.~\ref{Fig1_structure}(d)] are obtained by the opposite sign displacements according to the same pattern~\cite{Tan_arxiv2021,Christensen_arxiv2021}. Both SoD and iSoD structure display an in-plane $2 \times 2$ modulation of the high-temperature structure~\cite{Tan_arxiv2021}, which is clearly demonstrated by scanning tunneling microscopy (STM) measurements~\cite{Jiang2021NatureMaterial,Chen_arxiv2021,Zhao_arxiv2021,Liang_arxiv2021}.
Earlier x-ray diffraction measurements identified a three-dimensional $2\times2\times2$ CDW order in all AV$_3$Sb$_5$ compounds~\cite{Ortiz2020PhysRevLett, Li_arxiv2021}. However, more recent x-ray measurements reported a $2\times2\times4$ CDW order in CsV$_3$Sb$_5$~\cite{Ortiz_arxiv2021}, highlighting nontrivial interlayer ordering along the $c$-axis direction.
Furthermore, at an intermediate temperature $T^*$ about 60--70\,K, an additional uniaxial $1\times 4$ charge modulation was reported by STM studies~\cite{Chen_arxiv2021,Zhao_arxiv2021,Liang_arxiv2021}. Ultrafast coherent phonon spectroscopy measurements~\cite{Ratcliff_arxiv2021,Wang_arxiv2021,Wu_Kerr_arxiv2021}, muon spin relaxation measurements~\cite{Yu_arxiv2021usR}, and transport studies~\cite{Ni_arxiv2021,WenHH_arxiv2021} all identified anomalies at around $T^*$, suggesting a secondary instability below $T_S$.
Finally, density functional theory (DFT) phonon dispersion calculations found imaginary phonon frequencies at both $M (1/2,0,0)$ and $L (1/2,0,1/2)$ points~\cite{Tan_arxiv2021}, as well as at points along the $U$ line connecting the $M$ and $L$ momenta of the Brillouin zone for the high-temperature phase~\cite{Christensen_arxiv2021}, indicating $M^+_1$, $L^-_2$, and $U_1$ lattice instabilities. 
These $M^+_1$, $L^-_2$, and $U_1$ lattice instabilities contribute to $2\times2\times1$, $2\times2\times2$, and $2\times2\times4$ CDW orderings, respectively.
Which lattice instabilities give rise to the leading order parameter, and how it interplays with other lattice instabilities below the CDW transition in AV$_3$Sb$_5$ system, are not yet  conclusively understood~\cite{Christensen_arxiv2021}.

In this paper, we use polarization-resolved electronic Raman spectroscopy and DFT calculations to study the leading order parameter associated with the CDW transition in the kagome metal AV$_3$Sb$_5$.    
Prior x-ray diffraction data established that the CDW order in CsV$_3$Sb$_5$ has $2\times2\times4$ super-modulation structure with space group $P\bar{3}$: one layer of iSoD, alternating with three consecutive layers of SoD with zero-phase-shift between neighboring layers~\cite{Ortiz_arxiv2021}. 
To a good approximation, the $2\times2\times4$ structure can be further refined to space group $P6/mmm$, which is the same space group for the high temperature undistorted kagome phase. 
By analyzing the vanadium lattice distortions in the $2\times2\times4$ structure, we find that $U_1$ lattice instability is primary-like while $M_1^+$ and $L_2^-$ instabilities are secondary-like.
This is also confirmed by the calculation of the bare static susceptibility $\chi'_0(q)$ that shows a broad peak at around $q_z=0.25$ along the hexagonal Brillouin zone face central line ($U$ line).

These primary-like and secondary-like order parameters are revealed by Raman studies of the amplitude modes. 
We identify several new $A_{1g}$ and $E_{2g}$ phonon modes related to V and Sb atoms as well as alkali-metal atoms displacements in the CDW state. Both the new $A_{1g}$ and $E_{2g}$ modes are the amplitude modes of the CDW order parameter. Especially, each $A_{1g}$ new mode is a doublet that contains two modes close to each other. 
The detailed temperature evolution of the $A_{1g}$ modes' frequency, half-width at half-maximum, and integrated intensity support two successive phase transitions in CsV$_3$Sb$_5$: the first one with a primary-like order parameter appearing at $T_S=94$\,K and the second isostructural one emerging at about $T^*=70$\,K. 
Moreover,
we find two types of $T$ dependence of the integrated intensity for the new $A_{1g}$ modes below $T_S$: the low-energy modes show a plateau-like behavior below $T^*$ while the high-energy modes monotonically increase below $T_S$.
These two behaviors are captured by the Landau free-energy model incorporating the interplay between the primary-like and the secondary-like order parameters via trilinear coupling.

The rest of this paper is organized as follows. 
In Sec.~\ref{Results}, we present and discuss the theoretical and
experimental results.            
Specifically, in Sec.~\ref{Data_overview}, we first present an overview of the $T$ dependence of the Raman results.
In Sec.~\ref{Lattice_instabilities}, we  introduce three lattice instabilities along the $U$-line, namely, $M^+_1$, $L^-_2$, and $U_1$ instabilities.
In Sec.~\ref{Bare_susceptibility}, we examine the bare susceptibility $\chi'_0(q)$ along the $U$ line and establish an enhancement of $\chi'_0(q)$ at $q_z=0.25$, which corresponds to the $U_1$ lattice instability.
In Sec.~\ref{Primary_order}, we discuss the crystal structure of CsV$_3$Sb$_5$ at 15\,K and  analyze the major lattice distortions in CsV$_3$Sb$_5$.  
In Sec.~\ref{SUBDUCTION}, we perform the subduction analysis for the CDW phase.
In Sec.~\ref{Landau}, we construct a Landau free-energy model to study the interplay between the primary-like and secondary-like order parameters.
In Sec.~\ref{Phonon_modes}, we show several Raman modes appearing in the CDW state and compare them in three AV$_3$Sb$_5$ systems.   
In Sec.~\ref{Temperature_dependence}, we discuss the temperature dependence of the main phonon modes in $RR$ and $RL$ scattering geometries in CsV$_3$Sb$_5$.
In Sec.~\ref{New_modes}, we discuss the temperature dependence of the $A_{1g}$ and $E_{2g}$ amplitude modes in $RR$  and $RL$ scattering geometries in CsV$_3$Sb$_5$, respectively. 
Finally, in Sec.~\ref{Conclusion}, we provide a summary of our observations and conclusions.

\section{Experiment and Methods}\label{Experiment and Methods}

\subsection{Single crystal preparation and characterization}\label{Crystal_preparation}                                                                                                             
\begin{figure*}[!t] 
\begin{center}
\includegraphics[width=1.5\columnwidth]{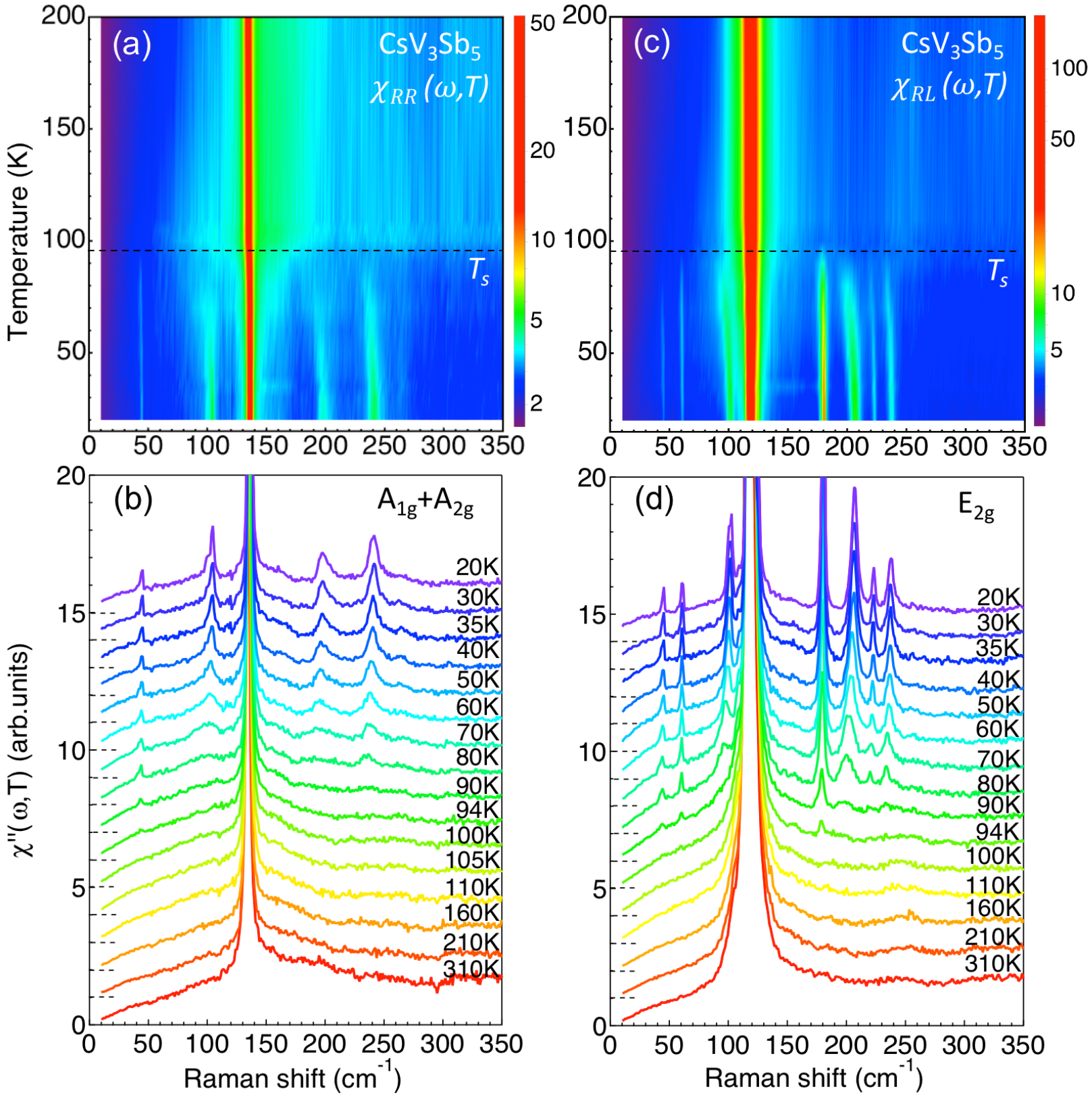}
\end{center}
\caption{\label{Fig3_T_dependence} 
(a) The color plot of the $T$-dependence of Raman spectra in $RR$ ($A_{1g}$+$A_{2g}$) scattering geometry for CsV$_3$Sb$_5$. (b) The corresponding Raman spectra of (a).
(c), (d) Same as (a) and (b) but for $RL$ ($E_{2g}$) scattering geometry. The dashed lines in (a) and (c) represent the structure phase transition temperature $T_S$. The dashed lines in (b) and (d) represent the level of zero for the vertically shifted spectra. 
}
\end{figure*}

Single crystals of AV$_3$Sb$_5$ (A= K, Rb, Cs) were synthesized via the flux method described in Refs.~\cite{Ortiz2019PhysRevMaterials,Ortiz2020PhysRevLett,Ortiz2021PhysRevMaterials}, and the chemical compositions
were determined by inductive coupled plasma
analysis. 
These samples were characterized by electric transport and magnetic susceptibility measurements.
The extracted structure phase transition (charge density wave transition) temperatures $T_{S}$ for AV$_3$Sb$_5$ (A= K, Rb, Cs) are 78, 103, and 94\,K, respectively~\cite{Ortiz2019PhysRevMaterials,Ortiz2020PhysRevLett,Ortiz2021PhysRevMaterials}. 
The sharpness of the Raman modes and the low residual spectra background (Fig.~\ref{Fig3_T_dependence}) indicate the high quality of the single crystals.

\subsection{Raman scattering measurements}\label{Raman}
                                             
The as-grown samples were cleaved in the air to expose a fresh (001) crystallographic plane.  
The fresh cleaved surface was stable in the air, as we did cleave the crystals in N$_2$ atmosphere and found no noticeable changes in the Raman data.
A strain-free area was examined by a Nomarski image. The cleaved crystals were positioned in
a continuous helium flow optical cryostat. The Raman measurements
were mainly performed using the Kr$^+$ laser line at 647.1\,nm (1.92\,eV) in
a quasibackscattering geometry along the crystallographic $c$ axis.
The excitation laser beam was focused into a $50\times100$ $\mu$m$^2$
spot on the $ab$ surface, with the incident power around 10\,mW. The
scattered light was collected and analyzed by a triple-stage Raman
spectrometer, and recorded using a liquid-nitrogen-cooled
charge-coupled detector. 
Linear and circular polarizations were used in this study to decompose the Raman data into different irreducible representations.
The instrumental resolution was maintained better than 1.5\,\cm-1. 
All linewidth data presented were corrected for the instrumental resolution. 
The temperatures were corrected for laser heating (Appendix~\ref{laser_heating_determination}).

All spectra shown were corrected for the spectral response of the spectrometer and charge-coupled detector to obtain the Raman intensity $I
_{\mu v}$, which is related to the Raman response $\chi''(\omega,T)$: $I_{\mu v}(\omega, T)=[1+n(\omega, T)] \chi_{\mu \nu}^{\prime \prime}(\omega, T)$. Here $\mu v$ denotes the polarization of 
incident (scattered) photon, $\omega$ is energy, $T$ is temperature, and $ n(\omega, T)$ is the Bose factor.
                                                                                                                                                                                        
The Raman spectra were recorded from the $ab$ (001) surface for scattering geometries denoted as $\mu v = XX, XY, RR, RL$, which is short for $Z(\mu v)\bar{Z}$ in Porto’s notation, where $X$ and $Y$ denote linear polarization parallel and perpendicular to the crystallographic axis, respectively; $R=X+iY$ and $L=X-iY$ denote the right- and left-circular polarizations, respectively. The $Z$ direction corresponds to the $c$ axis perpendicular to the (001) plane [see Fig.~\ref{Fig1_structure}(b)]. The polarization leakage 
from optical elements was removed in our data analysis (see Appendix~\ref{leakage}).

The relationship between the scattering geometries and the probed symmetry channels are summarized in Table~\ref{SymmetryAnalysis}. The algebra used in this study to decompose the Raman data into three irreducible representations of the point group $D_{6h}$ are summarized in Table~\ref{decompositionD6h}. More details are presented in Appendix~\ref{Raman_tensor_analysis}.
                                            
\begin{table}[t]
\caption{\label{SymmetryAnalysis}The relationship between the scattering geometries and the symmetry channels. $A_{1g}$,  $A_{2g}$, and $E_{2g}$ are the irreducible representations of the $D_{6h}$ point group.
}
\begin{ruledtabular}
\begin{tabular}{cc}
Scattering geometry&Symmetry channel\\
\hline
$XX$&$A_{1g}+E_{2g}$\\
$XY$&$A_{2g}+E_{2g}$\\
$RR$&$A_{1g}+A_{2g}$\\
$RL$&$2E_{2g}$\\
\end{tabular}
\end{ruledtabular}
\end{table}
                                       
\begin{table}[b]
\caption{\label{decompositionD6h}The algebra used in this study to decompose the Raman data into three irreducible representations of the point group $D_{6h}$.}
\begin{ruledtabular}
\begin{tabular}{cc}
Symmetry Channel&Expression\\
\hline
$A_{1g}$&$\chi''^{D_{6h}}_{XX}-\chi''^{D_{6h}}_{RL}/2$\\
$A_{2g}$&$\chi''^{D_{6h}}_{XY}-\chi''^{D_{6h}}_{RL}/2$\\
$E_{2g}$&$\chi''^{D_{6h}}_{RL}/2$\\
\end{tabular}
\end{ruledtabular}
\end{table}    
                        
\subsection{Density functional theory calculations}\label{DFT}
Density functional theory (DFT) calculations were performed within the Perdew-Burke-Ernzerhof-type generalized gradient approximation~\cite{Perdew1996PhysRevLett}, which is implemented in the Vienna $ab$-initio Simulation Package (VASP)~\cite{Kresse1996_PhysRevB,KRESSE199615}. The projected augmented wave potentials with 9 valence electrons for the A atom, 5 valence electrons for V, and 5 valence electrons for Sb were employed. The cutoff energy for the
plane-wave basis set was 300\,eV. The zero-damping DFT-D3 van der Waals correction was employed throughout the calculations. The phonon dispersion was calculated by using the finite displacement method as implemented
in the \uppercase{phonopy} code~\cite{TOGO20151}. More details of phonon calculations are presented in Ref.~\cite{Tan_arxiv2021}.

 \begin{figure*}[!ht] 
\begin{center}
\includegraphics[width=2\columnwidth]{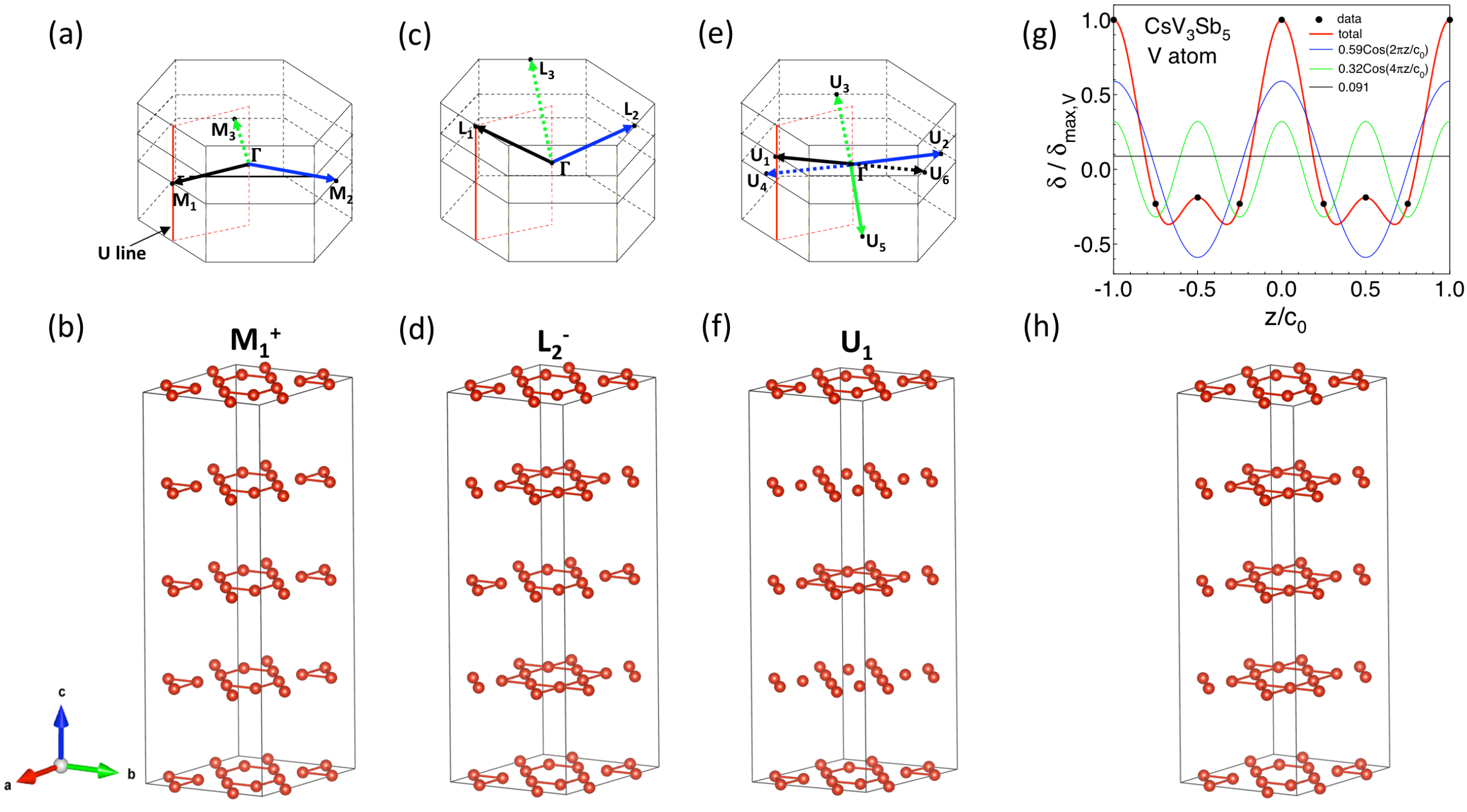}
\end{center}
\caption{\label{Fig2_BZ} 
The 3D hexagonal Brillouin zone (BZ) corresponds to the space group $P6/mmm$ with $M$, $L$, and $U$ points highlighted, as well as the $U$ line connecting $M$ and $L$ shown in red. 
(a) The three vectors in the stars of $M$ points, leading to $2\times2\times1$ CDW ordering shown as bond patterns in (b).
(c) The three vectors in the stars of $L$ points, contributing to $2\times2\times2$ CDW ordering shown as bond patterns in (d). 
(e) The six vectors in the stars of $U$ points, which give rise to $2\times2\times4$ CDW ordering shown as bond patterns in (f). 
For these vectors shown in (a), (c), and (e), the solid arrows are pointing to the front faces while the dashed arrows are pointing to the back faces.
(g) The normalized in-plane amplitude of V displacements $\delta/\delta_{max,V}$ (at $6j$ sites in the first kagome layer, $12n$ sites in the second kagome layer, and $6k$ sites in the third kagome layer) 
as a function of the normalized coordinate ($z/c_0$) for the $2\times2\times4$ structure of CsV$_3$Sb$_5$ at 15\,K. $c_0$ is the $c$-axis lattice constant for the four-layer structure. The solid red, blue, and green lines represent the total fitted curve, the $c_0$-cosinusoidal modulation component, and the $c_0/2$-cosinusoidal modulation component, respectively.
(h) Illustration of the $2\times2\times4$ structure for vanadium lattice~\cite{Ortiz_arxiv2021}. In (b), (d), (f), and (h), the Cs and Sb atoms are omitted for simplification.}
\end{figure*}

The bare charge susceptibility was calculated via Eqs.~\ref{equation_Rechi0} and \ref{equation_Imchi0} where both intraband and interband contributions are considered~\cite{Johannes2008PhysRevB}:     
\begin{equation}\label{equation_Rechi0}
\chi^{\prime}_0(\mathbf{q})=\lim _{\omega \rightarrow 0}\chi^{\prime}(\mathbf{q},\omega)\sim\sum_{\mathbf{k}} \frac{f\left(\varepsilon_{\mathbf{k}}\right)-f\left(\varepsilon_{\mathbf{k+q}}\right)}{\varepsilon_{k}-\varepsilon_{\mathbf{k}+\mathbf{q}}}
\end{equation}
\begin{equation}\label{equation_Imchi0}
\lim _{\omega \rightarrow 0} \chi^{\prime \prime}(\mathbf{q}, \omega) / \omega \sim \sum_{k} \delta\left(\varepsilon_{\mathbf{k}}-\varepsilon_{F}\right) \delta\left(\varepsilon_{\mathbf{k+q}}-\varepsilon_{F}\right).
\end{equation}
 $\chi_0^{\prime}(\mathbf{q})$ and $\lim _{\omega \rightarrow 0} \chi^{\prime \prime}(\mathbf{q}, \omega) / \omega$ are the real and imaginary parts of bare susceptibility, respectively. $f(\varepsilon)$ is the Fermi-Dirac distribution function. $\varepsilon_k$ is the band dispersion, $\varepsilon_F$ is the Fermi energy, and $\mathbf{q}=(q_x,q_y,q_z)$.     
The two parts for the high-temperature phase are calculated with a tight-binding Hamiltonian based on the maximally localized Wannier functions~\cite{MOSTOFI2008685}. The $k$ mesh for the Brillouin zone integral is $150\times150\times80$. The temperature in the Fermi-Dirac distribution  is about 116\,K ($\sim$10\,meV).   
For the imaginary-part integral, the delta functions were replaced with the Lorentzian functions. The full width at half maximum for the Lorentzian function is about 10\,meV.

\subsection{Symmetry analysis}\label{space_group}
The ISODISTORT tool~\cite{Campbell_wf5017}, which is part of the ISOTROPY software suite, was used to analyze the lattice distortions in CsV$_3$Sb$_5$ shown in Table~\ref{Instability}. This procedure uses projection operators that decompose lattice distortions into separate irreducible representations (irreps) of the space group. The amplitudes of these irreducible representations can then be analyzed to identify the primary-like lattice distortions that have the largest amplitude below the structural phase transition, and the secondary-like distortions which have smaller amplitudes.
We refrain from referring to these order parameters as ``primary'' and ``secondary'', and instead call them ``primary-like'' and ``secondary-like'', because there are trilinear terms present in the lattice Hamiltonian which can induce avalanche transitions. In such a phase transition, both order parameters set in together, but the transition disappears when the secondary-like order parameter is removed.
The free-energy expression in Eq.~\ref{free_energy} was obtained following the same procedure in Ref.~\cite{Christensen_arxiv2021}, and with the help of the INVARIANTS tool of the ISOTROPY software suite. The information for the irreducible representations of point groups and space groups follow the notations of Cracknell, Davies, Miller \& Love~\cite{cracknell1979general}, which is the same for the Bilbao Crystallographic Server~\cite{Bilbao_4,Bilbao_2}.

\section{Results and Discussions}\label{Results} 
     
\subsection{Data overview}\label{Data_overview}  
                      
\begin{figure*}[!t] 
\begin{center}
\includegraphics[width=1.8\columnwidth]{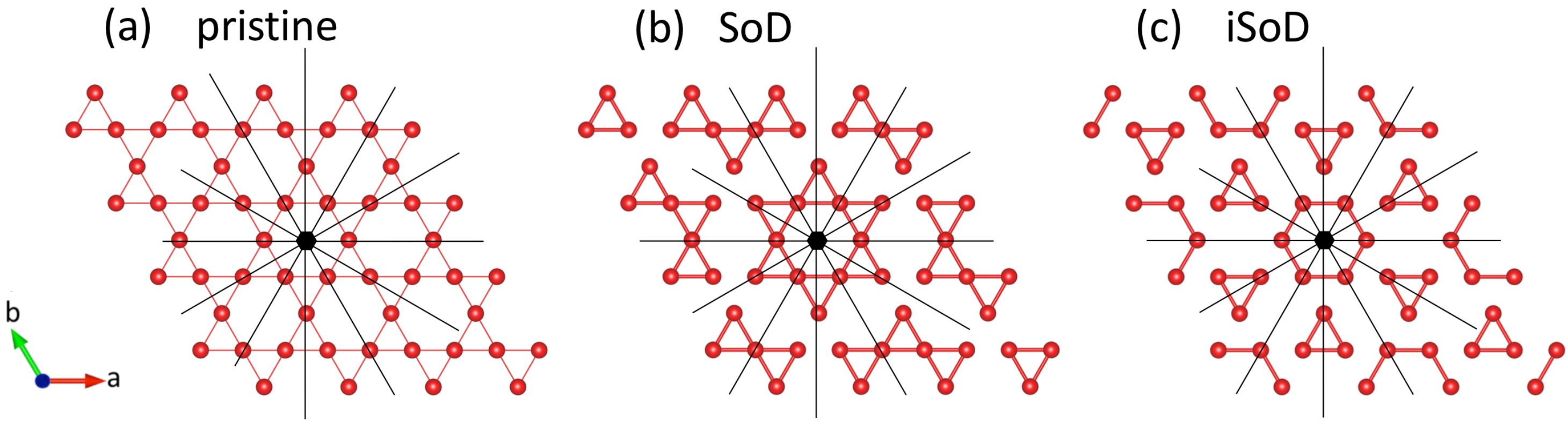}
\end{center}
\caption{\label{point_group_symmetry} 
Illustration of the $D_{6h}$ point group symmetries of the high-temperature, SoD, and iSoD phases of CsV$_3$Sb$_5$. The black lines represent the mirror planes. 
The $C_6$ rotational symmetry is preserved in the above three phases. The Sb and Cs atoms are omitted for simplification.}
\end{figure*}

The high-temperature undistorted crystal of CsV$_3$Sb$_5$ has a hexagonal structure with space group $P6/mmm$ (\#191) (point group:~$D_{6h}$) above $T_S=94$\,K. Below $T_S$, the translational symmetries are broken and a three dimensional CDW order forms. In Fig.~\ref{Fig3_T_dependence}, we present an overview of the $T$ dependence of the Raman modes in both $RR$ and $RL$ scattering geometries for CsV$_3$Sb$_5$. In addition to one main phonon in $RR$ scattering geometry and one main phonon in $RL$ scattering geometry persisting across $T_S$ upon cooling, several new phonons modes appear in both scattering geometries below $T_S$. 
The new modes in $RL$ scattering geometry and the higher-energy modes in $RR$ scattering geometry abruptly appear below $T_S$ and gain intensity gradually upon cooling [Fig.~\ref{Fig3_T_dependence}(c)]. In contrast, the low-energy modes in $RR$ scattering geometry first appear as relatively broad features which then sharpen upon cooling and gain intensity moderately~[Fig.~\ref{Fig3_T_dependence}(a)]. In the following sections, we will discuss the nature of these modes, analyze the leading order parameters, the symmetry of the CDW ground state, and the origin of these two different behaviors for the new Raman modes below $T_S$.

\subsection{Lattice instabilities at $M$, $L$, and $U$ points}\label{Lattice_instabilities}       
                                                                         
To begin with, we discuss the lattice instabilities in CsV$_3$Sb$_5$. 
Previous DFT calculations for this
crystal structure find two unstable phonon modes at $M$
and $L$ points, as well as at the middle points of the $U$-line connecting the $M$ and $L$ momenta of the Brillouin zone~\cite{Tan_arxiv2021,Christensen_arxiv2021} (see Appendix D).
They transform as $M_1^+(a, a, a)$, $L_2^-(a, a, a)$, and $U_1(a, -a; a, -a; a, -a)$ irreducible representations of the space group $P6/mmm$.      
Formally, $M_1^+(a, a, a)$, $L_2^-(a, a, a)$, and $U_1(a, -a; a, -a; a, -a)$ are one-dimensional irreducible representations of the little group of the wave vectors $M$($1/2,0,0$), $L$($1/2,0,1/2$), and $U$($1/2,0,1/4$), respectively. There are three vectors in the stars of both $M$ and $L$ points, as shown in Figs.~\ref{Fig2_BZ}(a) and (c), respectively.  The star of $U$ points has six vectors existing in three pairs. Within each pair, the two vectors are related to each other by inversion symmetry, as shown in Fig.~\ref{Fig2_BZ}(e). As a result, the $M_1^+$ and $L_2^-$ space group representations are three dimensional while the $U_1$ space group representation is six- dimensional.
                                                                                                                                                                                                                                  
For a single kagome layer, considering the $M_1^+$ instability, 
the equal-weight superposition of the three in-plane V-V displacements along the vectors shown in Fig.~\ref{Fig2_BZ}(a) lead to two distinct $C_6$-symmetric CDW orders.
If all three components of V-V displacements have the same phase and equal amplitudes [Fig.~\ref{Fig1_structure}(c)], the resulting shorter V-V bond pattern is the inverse star-of-David (iSoD) bond order shown in Fig.~\ref{Fig1_structure}(e). On the other hand, if we shift the phase of one of the three V-V displacements by $\pi$ while keeping the amplitudes the same [Fig.~\ref{Fig1_structure}(d)], the resulting pattern is the star-of-David (SoD) bond order shown in Fig.~\ref{Fig1_structure}(f). 
Flipping the sign of two components of the order parameter, in other words, shifting the phase of two of the three V-V displacements by $\pi$, does not change the CDW pattern, but it shifts the CDW pattern by one unit cell, leading to a different domain of the same low-symmetry phase.
While the translational symmetries are broken in the SoD and iSoD phases, they both possess the same set of mirror planes as the high-temperature structure, hence, they have the same point group $D_{6h}$, as is illustrated in Fig.~\ref{point_group_symmetry}.

In real space, the only difference for $M_1^+$, $L_2^-$, and $U_1$ instabilities is the interlayer ordering for neighboring V kagome layers along the $c$-axis direction. 
For $M_1^+$ instability, all $V$ atoms are displaced in phase between the neighboring layers [Fig.~\ref{Fig2_BZ}(b)], corresponding to a $2\times2\times1$ ordering. For $L_2^-$ instability, all V displacements are out-of-phase between the nearest-neighboring layers [Fig.~\ref{Fig2_BZ}(d)], leading to a $2\times2\times2$ ordering. 
For $U_1$ instability, all V displacements are out-of-phase between next-nearest-neighboring V layers [Fig.~\ref{Fig2_BZ}(f)], contributing to a $2\times2\times4$ ordering. 
Note that even when the $L_2^-$ and $U_1$ order parameters set in (when the phase of displacements in neighboring kagome layers are not equal), it is possible to have mirror planes both normal to and in the plane of kagome layers. The equal-amplitude distortions of all three $L_2^-$ or all six $U_1$ components give rise to the same $P6/mmm$ structures, but with enlarged ($2\times2\times2$ or $2\times2\times4$) unit cells. 
The realized CDW structure with $2\times2\times4$ supercell~[Fig.~\ref{Fig2_BZ}(h)] can be a combination of the three lattice instabilities, namely, it is a superimposition of the three bond orders shown in Figs.~\ref{Fig2_BZ}(b), (d), and (f). 
Because the $U_1$ has lower symmetries than $L_2^-$ or $M_1^+$, in other words, it breaks all the symmetries that $L_2^-$ and $M_1^+$ break, these $L_2^-$ and $M_1^+$ order parameters whose displacement patterns will be induced by $U_1$ can be regarded as secondary(-like) order parameters.

\begin{figure*}[!t] 
\begin{center}
\includegraphics[width=2\columnwidth]{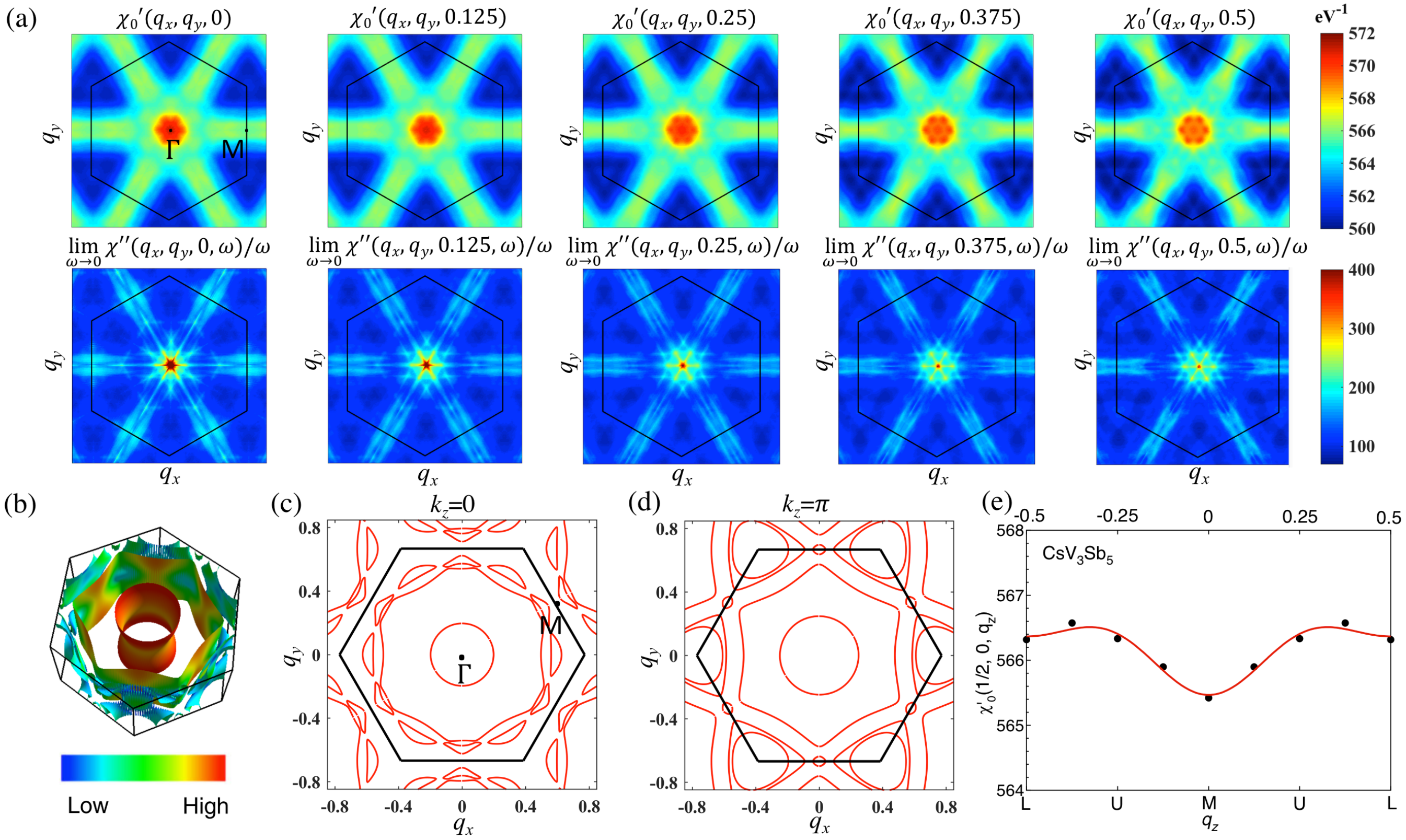}
\end{center}
\caption{\label{CVS_Chi0} 
(a) The bare susceptibility $\chi'_0(q_x,q_x,q_z)$ and $\lim _{\omega \rightarrow 0} \chi^{\prime \prime}(q_x,q_y,q_y, \omega) / \omega$ for CsV$_3$Sb$_5$ in the high-temperature phase for $q_z$=0, 0.125, 0.25, 0.375, and 0.5, respectively. 
(b) Three-dimensional Fermi surfaces for CsV$_3$Sb$_5$ adapted from Ref.~\cite{Tan_arxiv2021}. 
(c) Two-dimensional Fermi-surface cut for CsV$_3$Sb$_5$ at $k_z=0$.
(d) Same as (c) but for $k_z=\pi$.
(e) $\chi'_0(1/2,0,q_z)$ along the $L-U-M-U-L$ line as a function of $q_z$. The solid red line is the guide line.
}
\end{figure*}    
                                                                                                                                                                                                                                                                                                  
\subsection{Bare susceptibility}\label{Bare_susceptibility}                                                                                  

In order to understand which lattice instabilities ($M_1^+$, $L_2^-$, and $U_1$) drive the CDW order in AV$_3$Sb$_5$, we examine the bare susceptibility $\chi_0(q_x,q_y,q_z)$.

In Fig.~\ref{CVS_Chi0}(a), we show $\chi'_0(q_x,q_y,q_z)$ and 
$\lim _{\omega \rightarrow 0} [\chi^{\prime \prime}(q_x,q_y,q_y, \omega) / \omega]$ for CsV$_3$Sb$_5$ in the high-temperature phase with $q_z=0, 0.125, 0.25, 0.375, 0.5$ at 116\,K. The $M^+_1$, $U_1$, and $L_{2}^{-}$ lattice instabilities locate in the $q_z=0, 0.25, 0.5$ planes, respectively. 
For $q_z=0$, $\chi'_0(q_x, q_y, 0)$ shows a broad enhancement along $\Gamma M$ direction in the momentum space, indicating that Fermi surface nesting is marginal to drive the CDW transition. 
This result is also consistent with the Ref.~\cite{Wang2022PhysRevB}.
Local minima are found at $M$ points and ridges along $\Gamma M$ direction exist around $M$ point. Similar topography in $\chi'_0(q_x, q_y, q_z)$ is also found for nonzero $q_z=0.125, 0.25, 0.375, 0.5$, as well as for $\lim _{\omega \rightarrow 0} \chi^{\prime \prime}(q_x,q_y,q_y, \omega) / \omega$. The ridges around $\Gamma M$ lines originate from the 2D-like large hexagonal Fermi surface of V $3d$ band near Brillouin zone boundary~[Fig.~\ref{CVS_Chi0}(b)]. Because the faces of the hexagon are relatively flat~[Fig.~\ref{CVS_Chi0}(c) and (d)], the susceptibility along the $\Gamma M$ direction parallel to the faces is large, resulting in the ridge along the $\Gamma M$ direction. Due to the 2D-like Fermi surface~[Fig.~\ref{CVS_Chi0}(b)], $\chi'_0(q_x,q_y,q_z)$ and $\lim _{\omega \rightarrow 0} \chi^{\prime \prime}(q_x,q_y,q_y, \omega) / \omega$ show little dependence of $q_z$~[Fig.~\ref{CVS_Chi0}(a)].

In Fig.~\ref{CVS_Chi0}(e), we present $\chi'_0(1/2,0,q_z)$ as a function of $q_z$ along the $U$ line, namely, $L-U-M-U-L$ line. 
$\chi'_0(1/2,0,q_z)$ show a general enhancement along the whole $U$-line compared with BZ corner. 
This indicates the $L-U-M-U-L$ line is a line of lattice instabilities, which is the source for nontrivial interlayer orderings along the $c$-axis direction~\cite{Ortiz2020PhysRevLett, Li_arxiv2021,Ortiz_arxiv2021}.
In particular, $\chi'_0(1/2,0,q_z)$ shows a broad peak at around $|q_z|=0.25$ along the $U$ line, which suggests that $U_1$ lattice instability is dominant compared with $M_1^+$ and $L_2^-$ lattice instabilities. 

Furthermore, due to the broad $q$ response, the calculated bare susceptibility $\chi'_0(q_x,q_y,q_z)$ does not indicate a single CDW ordering wave vector. 
The same is true also for the DFT phonon calculations~[Appendix~\ref{Phonon_instabilities}].
Therefore, x-ray diffraction refinement data are required to settle the CDW ordering wave vector as well as the dominant lattice instability symmetry. 
These results will be presented in the next section.
                                                                                                                                                                                                                                                                                               
We note that the true CDW instabilities can only be found from the full susceptibilities $\chi$, not merely from the bare susceptibility $\chi_0$. 
The full susceptibilities must account for both the electronic interactions and the interactions with the lattice. 
In addition, the local field effects might need to be considered in the calculations~\cite{Wang2022PhysRevB,Johannes2006_PhysRevB}.

\subsection{x-ray diffraction analysis}\label{x-ray} 
                             
\subsubsection{Primary-like and secondary-like order parameters}\label{Primary_order}
                                                                                                                                                        
\begin{figure*}[!t] 
\begin{center}
\includegraphics[width=1.8\columnwidth]{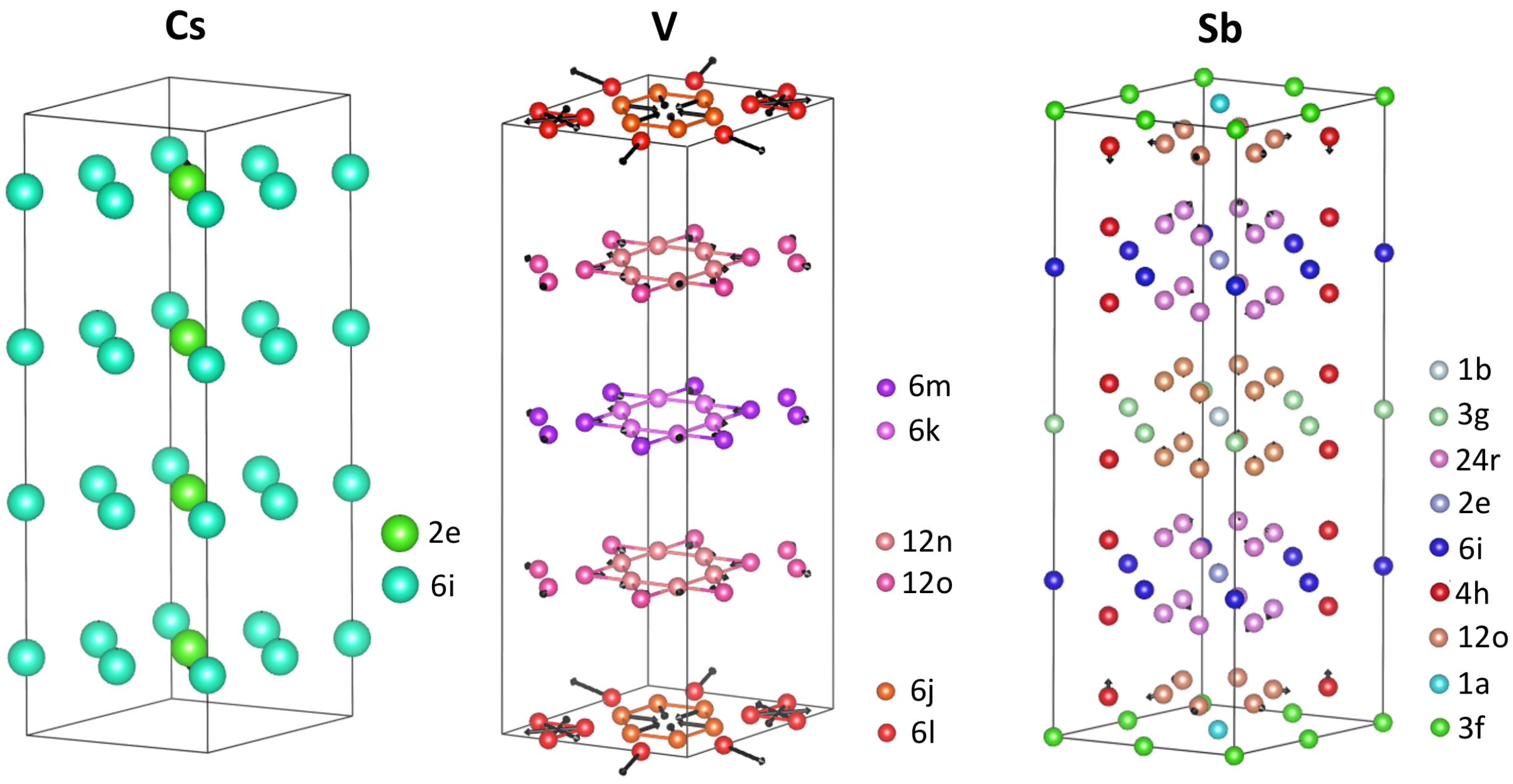}
\end{center}
\caption{\label{Fig6_displacement_pattern} 
Wyckoff-site dependent displacement pattern for CsV$_3$Sb$_5$ in the $2\times2\times4$ CDW phase at 15\,K~\cite{Ortiz_arxiv2021}. The Wyckoff sites are indexed according to the space group $P6/mmm$ based on the unit cell of the $2\times2\times4$ CDW phase.
The arrows represent the direction of displacement for specific atoms at 15\,K. The length of arrows for different atoms are scaled for visualization purpose.
}
\end{figure*}                                     
                                                                                                                            
For CsV$_3$Sb$_5$, 
the CDW order is established to be a $2\times2\times4$ structure below $T_S$=94\,K~\cite{Ortiz_arxiv2021}. It is refined within the space group $P\bar{3}$ assuming a minimal three-fold rotational symmetry and the inversion symmetry based on the x-ray diffraction data~\cite{Ortiz_arxiv2021}.                                                                                                                                                                                                                                                          
                                                    
In Fig.~\ref{Fig6_displacement_pattern}, we show the refined $2\times2\times4$ structure for CsV$_3$Sb$_5$ at 15\,K. It is composed of one layer of iSoD structure and three consecutive layers of SoD structures, with zero-phase-shift between neighboring layers. If we disregard the tiny distortions, $2\times2\times4$ structure can be further refined to higher-symmetry space group $P6/mmm$. 
The $D_{6h}$ point group symmetries of the high-temperature phase are also preserved in this $2\times2\times4$ $P6/mmm$ structure, similar as the SoD and iSoD structure illustrated in Fig.~\ref{point_group_symmetry}.
                                                   
To visualize the lattice distortion in the CDW phase, in Fig.~\ref{Fig6_displacement_pattern}, we show the displacement pattern for the $2\times2\times4$ $P6/mmm$ structure of CsV$_3$Sb$_5$ at 15\,K. The most prominent distortions come from V atoms in the bottom kagome layer at $6l$ and $6j$ Wyckoff sites, which show iSoD-type distortions. These distortions are about 1\% of the lattice constant, consistent with the 3$\sim$4 orders of magnitude weaker for the superlattice Bragg peak intensity compared with the fundamental lattice Bragg peak.
The V atoms in the other three kagome layers at $6m$, $6k$, $12n$, and $12o$ Wyckoff sites show SoD-type distortions, which are about four times smaller than the iSoD layer. For Sb atoms, only the ones in bottom honeycomb layer at $4h$ and $12o$ Wyckoff sites show noticeable distortions, the rest ones in the other three layers hardly move. For Cs atoms, the distortions at $2e$ Wyckoff sites are tiny while the ones at $6i$ Wyckoff sites barely move in the CDW phase.
                                                                                                                
\begin{table}[b]
\caption{\label{Instability} Analysis of major distortions in the refined $2\times2\times4$ $P6/mmm$ structure at 15\,K for CsV$_3$Sb$_5$ based on data in Ref.~\cite{Ortiz_arxiv2021}. The units of the amplitude are \AA. }
\begin{ruledtabular}
\begin{tabular}{ccc}
Displacement &Instability&Amplitude\\
\hline
Sb&$\Gamma^+_1$&0.13\\
\hline
V& $M^+_1(a, a, a)$&0.05\\
V& $L^-_2(a, a, a)$&0.17\\
V& $U_1 (a, -a; a, -a; a, -a)$&0.22\\
\hline
Cs& $L^-_2(a, a, a)$&0.03\\
 Cs&$U_1 (a, -a; a, -a; a, -a)$&-0.03\\
\end{tabular}
\end{ruledtabular}
\end{table}  
                                                                                                                                              
To quantitatively analyze these lattice distortions and figure out the leading order parameters, we use the ISODISTORT tool set~\cite{Campbell_wf5017} to list all the major distortions in Table~\ref{Instability}.                                                             
From Table~\ref{Instability}, for V displacements, the $U_1 (a, -a; a, -a; a, -a)$ distortion is the largest with an amplitude of 0.22~\AA. This is the total displacement, summed over all V atoms in the $2\times2\times4$ $P6/mmm$ supercell.
The $L^-_2(a, a, a)$ distortion is a bit smaller. The amplitude for $M^+_1(a, a, a)$ distortion is about 0.05~\AA, much smaller than the $U_1 (a, -a; a, -a; a, -a)$ distortion. 
Since the $U_1$ distortion is largest and it breaks all the in-plane and $c$-axis translational symmetries that are broken by $L^-_2$ and $M^+_1$ order parameters, we refer to it as the primary-like order parameter while we refer to the $L^-_2(a, a, a)$ or $M^+_1(a, a, a)$ distortions as secondary-like order parameters~\footnote{
The reason that we do not regard the $U_1$ distortion as the primary order parameter is due to the trilinear coupling between $U_1$,  $L^-_2$, and $M^+_1$ order parameters. This coupling leads to a simultaneous condensation of multiple order parameters at a temperature where all the second-order coefficients in the free-energy expansion become positive~\cite{Christensen_arxiv2021,Etxebarria2010}. This is referred to as an ``avalanche-transition''. Since the avalanche transition is a first-order transition, it is impossible to separate the primary and secondary order parameters as in an ordinary first-order phase transition. Hence, we refer to the order parameters only as primarylike and secondarylike.}.

Similar conclusions can be also reached by analyzing the normalized amplitude of V displacements in the four kagome layers. In Fig.~\ref{Fig2_BZ}(g), we present the normalized in-plane amplitude of V-displacements ($\delta/\delta_{max,\text{V}}$) (at $6j$ sites in the first kagome layer, $12n$ sites in the second kagome layer, $6k$ sites in the third kagome layer) as function of $z/c_0$ coordinate ($c_0$ is the $c$-axis lattice constant for the four-layer unit cell) at 15\,K. The positive sign of $\delta/\delta_{max,\text{V}}$ represents the iSoD-type distortion while the negative sign represents the SoD-type distortion. We find that $\delta/\delta_{max,\text{V}}$ can be modeled by a sum of three components: a period of $c_0$ interlayer cosinusoidal modulation, a period of $c_0/2$ interlayer cosinusoidal modulation, and a constant. These three components correspond to $U_1$ [Fig.~\ref{Fig2_BZ}(f)], $L^-_2$  [Fig.~\ref{Fig2_BZ}(d)], and $M^+_1$ [Fig.~\ref{Fig2_BZ}(b)] order parameters, respectively. The amplitude of $c_0$ interlayer modulation is about twice of $c_0/2$ interlayer modulation, consistent with previous distortion analysis for V displacements that $U_1$ order parameter is the dominant one that drives the $2\times2\times4$ CDW order while $L^-_2$ order parameter is secondary-like. Similar conclusions can also be reached by analyzing V displacements at $6l$ sites in the first kagome layer, $12o$ sites in the second kagome layer, and $6m$ sites in the third kagome layer.
                                                                                                                                               
\subsubsection{The symmetry of the order parameters}\label{SUBDUCTION}  
             
After establishing that $U_1$ instability is primary-like while $L^-_2$ and $M^+_1$ instabilities are secondary-like for V displacements, in this section we focus on the symmetry of the order parameters in the CDW ground state~\cite{Holy1976PRL}.

We refer to the high-temperature structure's space group as $G_{1\times1\times1}$. 
Similarly, we refer to the space group of the $2 \times 2 \times 4$ supercell as $G_{2\times2\times4}$.
Each symmetry of $G_{2\times2\times4}$ is a symmetry of $G_{1\times1\times1}$, but the opposite is not true, because the translation by a primitive lattice vector is no longer a symmetry in the $2\times2\times4$ structure. 
When the symmetry is reduced from $G_{1\times1\times1}$ to $G_{2\times2\times4}$ due to the reduced translational symmetry, the irreducible representations of $G_{1\times1\times1}$ may become reducible representations of $G_{2\times2\times4}$. 
The originally unstable $M_{1}^{+}$, $L_{2}^{-}$ and $U_1$ modes of the high-temperature structure get folded back to the center of the Brillouin zone for the low-temperature structure. 
The list of the irreducible representations subducted  from the irreducible representations of $G_{1\times1\times1}$ for $G_{2\times2\times4}$ are:     
\begin{equation}\label{equation1}
M_{1}^{+} {\downarrow} G_{2\times2\times4} =\Gamma_{1}^{+}+\Gamma_{5}^{+} \\
\end{equation}           
\begin{equation}\label{equation2}
L_{2}^{-} \downarrow G_{2\times2\times4}=\Gamma_{1}^{+}+\Gamma_{5}^{+} \\
\end{equation}
\begin{equation}\label{equation3}
U_{1} \downarrow G_{2\times2\times4}=\Gamma_{1}^{+}+\Gamma_{5}^{+}+\Gamma_{2}^{-}+\Gamma_{5}^{-}
 \end{equation}
Equations~(\ref{equation1})-(\ref{equation3}) show the symmetry-allowed order parameters with specific irreducible representation symmetries that can emerge in the ordered state due to the corresponding $M_{1}^{+}$, $L_{2}^{-}$, and $U_{1}$ instabilities.
For example,  $U_{1}$ instability leads to four potential order parameters in the $2\times2\times4$ ground state:  $\Gamma^+_1$ (which corresponds to the point group irreducible representation $A_{1g}$, Raman active, $z^2$ or $x^2+y^2$), $\Gamma^+_5$ (which corresponds to $E_{2g}$, Raman active, $x^2-y^2$ or $xy$), $ \Gamma^-_2$ (which corresponds to $A_{2u}$, infrared active, $z$), and $\Gamma^-_5$ (which corresponds to $E_{2u}$).

The $\Gamma^+_1$ ($A_{1g}$) is the fully symmetric irreducible representation and $\Gamma^+_5$ ($E_{2g}$) is the two-dimensional (2D) irreducible representations. Both representations are Raman active~\cite{NAGAOSA1982809}.
According to Eq.~(\ref{equation3}), only a single unstable phonon in the high-temperature structure freezes below the phase-transition temperature, and becomes the order parameter with a nonzero expectation value in the ground state.
This realizes the global minimum of the free energy, and transforms as $\Gamma^+_1$($A_{1g}$) in the low-temperature structure.

The other unstable modes in Eq.~(\ref{equation3}) can be thought of as the ``failed order parameters" that would appear as low-energy oscillations of the order parameter. 
The excitations into the ``failed order parameters" could be accessed spectroscopically by exciting the amplitude modes of the ``failed order parameters".

$\Gamma^+_5$ ($E_{2g}$), one of the ``failed order parameters", breaks the three-fold rotational symmetry. This ``nematiclike" $C_3$-symmetry broken-phase could be stabilized on the surface, which is predicted for a real CDW ordering with $k_z\neq 0$ due to the three-dimensional coupling~\cite{Park_arxiv2021}. 
It might explain the rotational symmetry breaking due to uniaxial charge modulation observed by STM~\cite{Liang_arxiv2021,Zhao_arxiv2021,Chen_arxiv2021}, two-fold $c$-axis magnetoresistance below $T_S$~\cite{Ni_arxiv2021,WenHH_arxiv2021}, the enhanced $m_{11}-m_{12}$ elasto-resistance coefficient above $T_S$~\cite{ZhenyuWang2021}, as well as the two-fold-like amplitude of coherent phonon mode observed in the pump-probe Kerr rotation measurement~\cite{Wu_Kerr_arxiv2021}.

\begin{figure*}[!t] 
\begin{center}
\includegraphics[width=2\columnwidth]{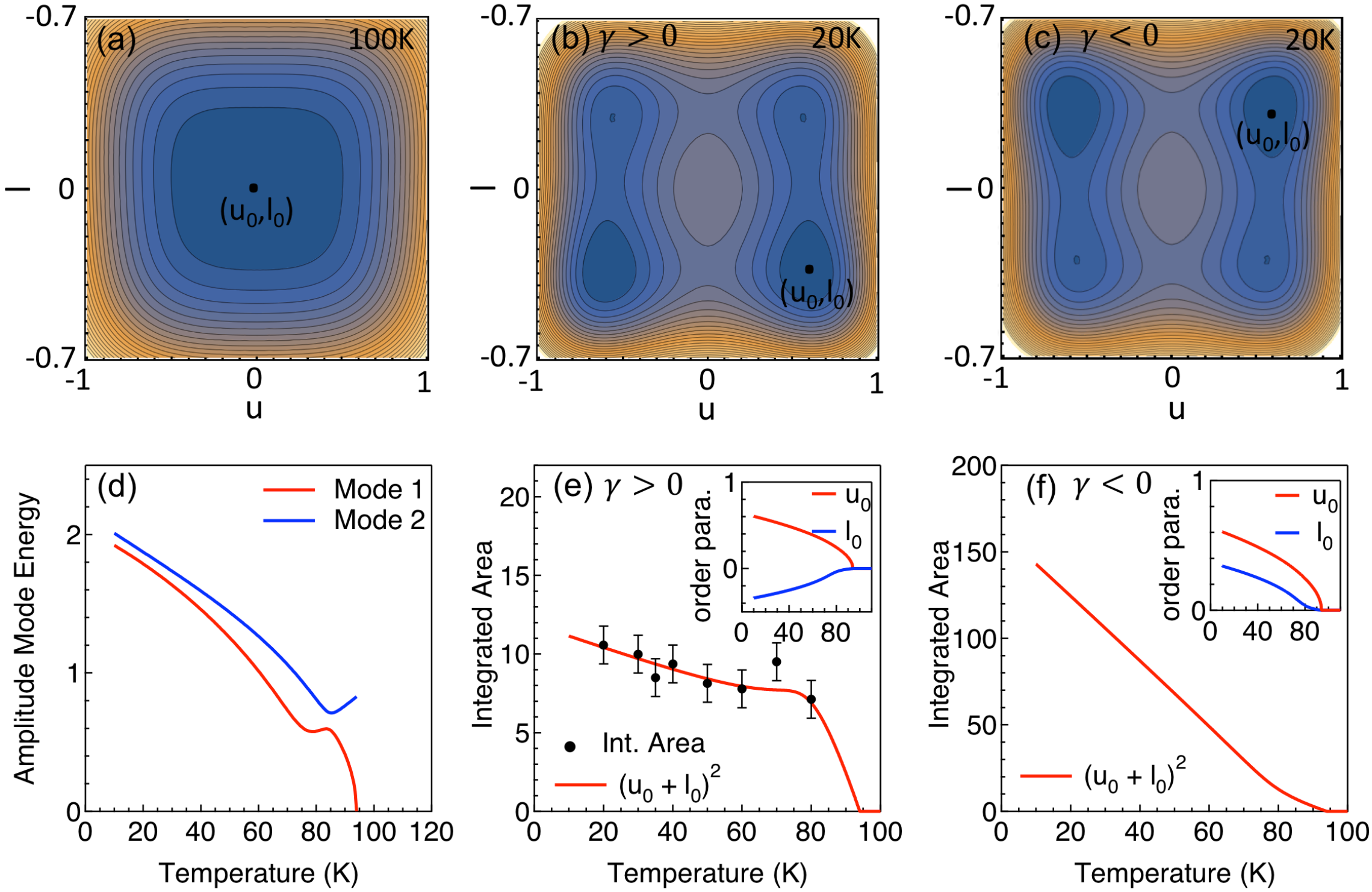}
\end{center}
\caption{\label{Order_parameter} 
Illustration of the Landau free-energy model (Eq.~\ref{free_energy}). The parameters used in the model are $\alpha_{u0}=-1$, $\alpha_{l0}=-1$, $\gamma=0.2$, $\beta=0.1$, $\lambda_u=1.3$, $\lambda_l=4$, $T_S=94$, and $T^*=70$.
(a) The free energy at 100\,K. 
(b) The free energy at 20\,K. 
(c) The free energy at 20\,K in the case of $\gamma=-0.2$ while the other parameters remain unchanged.
In (a)-(c), the global minima are represented by a black dot at $(u_0,l_0)$.
(d) $T$ dependence of the $A_{1g}$ amplitude mode frequencies for mode 1 and 2.
(e) $T$ dependence of $(u_0+l_0)^2$. The black solid circles represent the $T$-dependence of the integrated intensity for the $A_{1g}$ 105\,\cm-1 mode in CsV$_3$Sb$_5$. The inset of (e) shows the $T$ dependence of $u_0(T)$ and $l_0(T)$ order parameters.
(f) $T$ dependence of $(u_0+l_0)^2$ in the case of $\gamma=-0.2$ while the other parameters remain unchanged. The inset of (f) shows the $T$ dependence of $u_0(T)$ and $l_0(T)$ order parameters.
}
\end{figure*} 
                                                                                                                                                                                                               
\subsubsection{Interplay between the primary-like and secondary-like order parameters}\label{Landau}                                                                                            
                                                                              
In this section, we use a phenomenological model at the level of Landau free-energy expansion to study the interplay between the primary-like and secondary-like order parameters for V displacements in the ground state with $\Gamma^+_1$ ($A_{1g}$) symmetry.
                                                                                                                                                                                   
Since the amplitude of the secondary-like $M^+_1$ distortion is significantly smaller than the $L^-_2$ and $U_1$ distortions of V displacements~(Table~\ref{Instability}), we neglect $M^+_1$ distortion. For the specific domain and order parameter directions of $L^-_2$ and $U_1$, the free energy is 
\begin{equation}\label{free_energy}
\mathcal{F}(T)=\alpha_{u}(T) u^{2}+\alpha_{l}(T) l^{2}+\gamma u^{2}l+\beta u^{2} l^{2}+\lambda_{u} u^{4}+\lambda_{l} l^{4},   
\end{equation} 
where $u$ and $l$ represent the $U_1$ and $L^-_2$ order parameters, respectively. $\alpha_{u}$, $\alpha_{l}$, $\lambda_u$, and $\lambda_l$ are coefficients while $\gamma$ and $\beta$ are coupling constants between $u$ and $l$. 
This form of free energy is found by considering every polynomial, up to fourth order, of the order parameters that remain invariant under each symmetry operation of the high-temperature space group. 
The $\gamma u^{2}l$ term is the trilinear term while the $\beta u^{2} l^{2}$ terms is the biquadratic coupling term. They involve coupling between different order parameters~\cite{Christensen_arxiv2021}.

According to DFT phonon calculations which are performed at zero temperature~\cite{Tan_arxiv2021,Christensen_arxiv2021}, both the $U_1$ and $L^-_2$ modes are unstable. Therefore, both $\alpha_{u}$ and $\alpha_{l}$ are negative at zero temperature; however, at high temperature they must both be positive so that the non-symmetry-broken structure is retained. There is no symmetry argument that sets the temperature at which these parameters cross zero, so one can set $\alpha_{u}(T)=\alpha_{u0}(T-T_S)$ and $\alpha_{l}(T)=\alpha_{l0}(T-T^*)$ without loss of generality. Doing this, and minimizing the free energy Eq.~\ref{free_energy} with respect to $u$ and $l$, we can obtain the solution of $u_0(T)$ and $l_0(T)$, as well as the corresponding free energy $\mathcal{F}(T)$.

As shown in Fig.~\ref{Order_parameter}(a), well above $T_S$ the free energy has a single minimum at $(0,0)$. Reducing the temperature near $T_S$, nonzero order parameters $u_0$ and $l_0$ develop. The free energy develops two minima in the parameter space.
Depending on the sign of the parameter $\gamma$, the global minimum appears in a different part of the parameter space. Suppose a case that $\gamma>0$ and the global minimum appears where $u_0>0$ and $l_0<0$~[Fig.~\ref{Order_parameter}(b)]. If we reverse the sign of $\gamma$, the global minimum locates where $u_0>0$ and $l_0>0$~[Fig.~\ref{Order_parameter}(c)]. This conclusion also holds when $u<0$ such as in a domain structure, because $\mathcal{F}(T)$ is even with respect to $u$.
In the insets of Figs.~\ref{Order_parameter}(e) and (f), we show the solution $u_0(T)$ and $l_0(T)$ corresponding to the global minimum in Figs.~\ref{Order_parameter}(b) and (c), respectively. 
In both cases, the transition is first order and both order parameters have a discontinuity at the transition temperature, but the primary-like order parameter $u_0(T)$ develops more sharply below $T_S$ while the secondary-like order parameter $l_0(T)$ develops  gradually below $T_S$.   
                                                                                                                
The $T$ dependence of the square of the order parameters $(u_0+l_0)^2$ can distinguish how the primary-like and secondary-like order parameters interplay with each other. In Fig.~\ref{Order_parameter}(e), we show the $T$ dependence of $(u_0+l_0)^2$ in the case of $\gamma>0$ where $u_0$ and $l_0$ have different signs. In this case, $u_0$ and $l_0$ interplay with each other destructively. As a consequence, $(u_0+l_0)^2$ shows a plateau-like behavior below $T^*$.
In Fig.~\ref{Order_parameter}(f), we show the $T$ dependence of the $(u_0+l_0)^2$ in the case of $\gamma<0$ where $u_0$ and $l_0$ have the same signs. In this case, $u_0$ and $l_0$ interplay with each other constructively. Thus, 
$(u_0+l_0)^2$ increases monotonically below $T_S$.       
These two behaviors can be tested by the temperature dependence of the amplitude modes' integrated intensities, which will be discussed in Section.~\ref{New_modes}.

Moreover, the $T$ dependence of $(u_0+l_0)^2$ also implies the properties of the second phase transition at $T^*$. For V displacements, the first transition at $T_S$ is driven by the primary-like $U_1$ order parameter, and it breaks both the in-plane and $c$-axis translational symmetries at $T_S$. 
The secondary-like order parameter $L^-_2$ appears by coupling to the primary-like order parameter below $T_S$.
The second transition involves mostly a change in the secondary-like order parameter's amplitude at $T^*$. 
This second transition is isostructural, and results in no change in the symmetry of the crystal. 
As a consequence, this second transition at $T^*$ is necessarily first order according to group theory.

Finally, we discuss the $T$ dependence of the $A_{1g}$ amplitude mode's frequencies. 
We expand the free energy $\mathcal{F}(T)$ at the minimum position ($u_0$, $l_0$).
Taking the second derivative of $\mathcal{F}(T)$ with respect to $u$ and $l$,
and solving the eigenvalue of the equation
\begin{align}\label{amplitude_mode}
\begin{vmatrix}
\partial^{2} \mathcal{F}/\partial u^{2}-m \omega^2 & \partial^{2} \mathcal{F}/\partial u \partial l  \\ 
\partial^{2} \mathcal{F}/\partial l \partial u & \partial^{2} \mathcal{F}/\partial l^{2}-m \omega^2  \\ 
\end{vmatrix}
_{u=u_0(T), l=l_0(T)}=0,
\end{align}                 
we obtain the $T$ dependencies of two normal $A_{1g}$ amplitude mode frequencies $\omega$ around the free-energy minima at ($u_0$, $l_0$)~\footnote{In Eq.~(\ref{amplitude_mode}), $m$ has the unit of the mass in analog of a two-dimensional harmonic oscillator.}. 
We note that there are two solutions of $A_{1g}$ amplitude mode frequencies based on Eq.~(\ref{amplitude_mode}), suggesting that each $A_{1g}$ amplitude mode is a doublet that contains two modes close to each other in this system.
                                                               
In Fig.~\ref{Order_parameter}(d), we show an example of the $T$ dependence of the two amplitudes modes frequencies below $T_S$.  Mode 1 increases gradually from zero below $T_S$. Mode 2 first decreases, showing an upturn, and then increases gradually. 
Both modes 1 and 2 show a clear anomaly at around 80\,K, which is close to $T^*$ that is set to be 70 in the free-energy model. 
The anomalies at around $T^*$ are due to the interplay between the primary-like and secondary-like order parameters, because the frequencies of mode1 and 2 are expected to show mean-field-like behavior (square root of $T_S-T$) in the case of $\gamma=0$ and $\beta=0$. 
                                                                                    
\subsection{Raman results} \label{Raman_results}     

In the previous sections, we have established the primarylike and secondarylike order parameters for V displacements, and the interplay between them in the free-energy model. In this section, we illustrate how they are reflected in the Raman data.
                                                                                             
\subsubsection{Phonon modes} \label{Phonon_modes}                    

\begin{figure*}[!t] 
\begin{center}
\includegraphics[width=1.7\columnwidth]{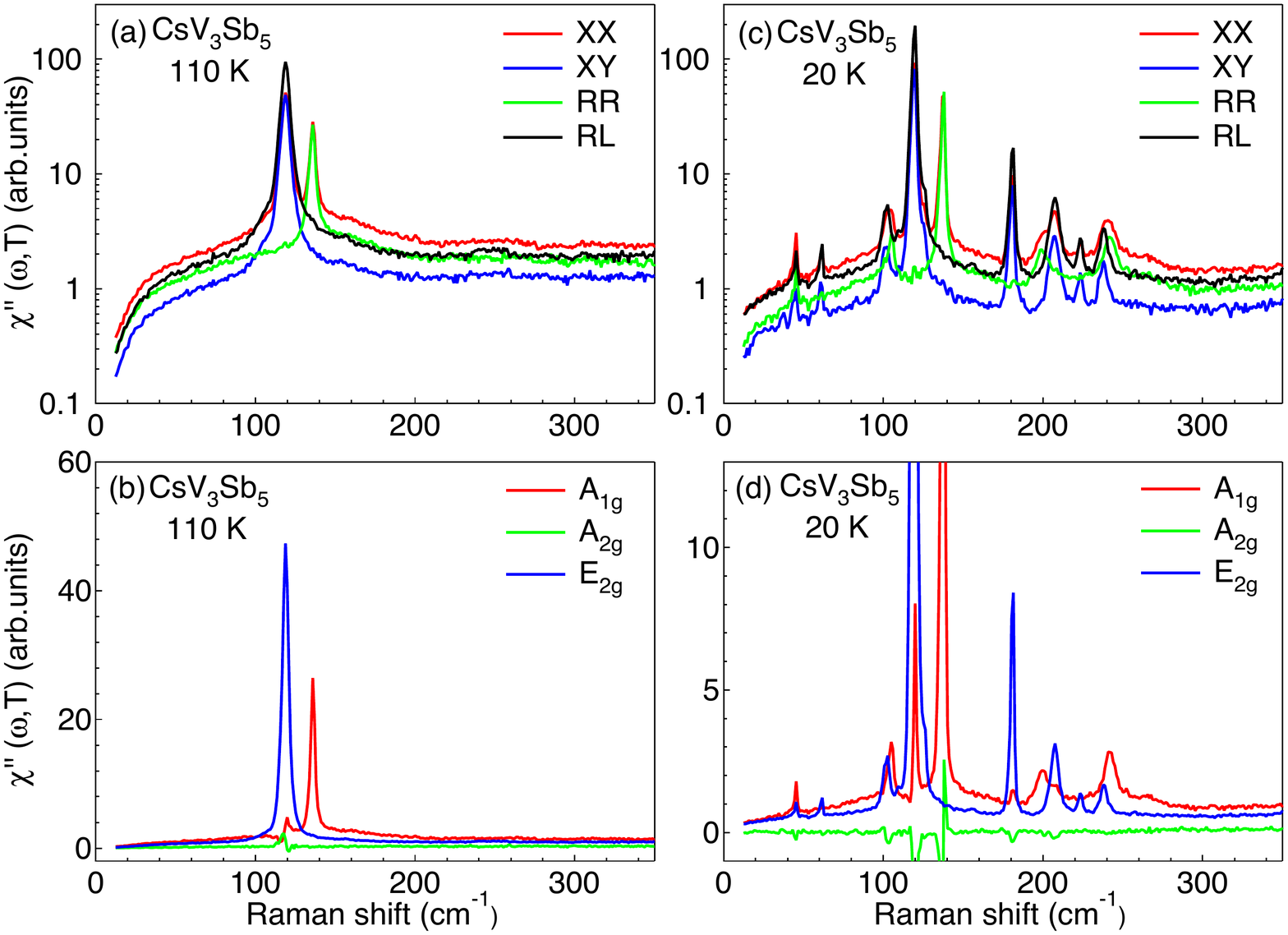}
\end{center}
\caption{\label{Fig2_CDW_phonon}
Symmetry-resolved spectra of  CsV$_3$Sb$_5$ above and below $T_S$. (a) Raman spectra of CsV$_3$Sb$_5$ on a cleaved $ab$ surface for the $XX$, $XY$, $RR$, and $RL$ scattering geometries at 110\,K.  (b) Symmetry decompositions into separate irreducible representations according to the point group $D_{6h}$ using the algebra shown in Table~\ref{decompositionD6h}. 
(c), (d) Same as (a) and (b) but at 20\,K.
}
\end{figure*}

\begin{figure}[!t] 
\begin{center}
\includegraphics[width=\columnwidth]{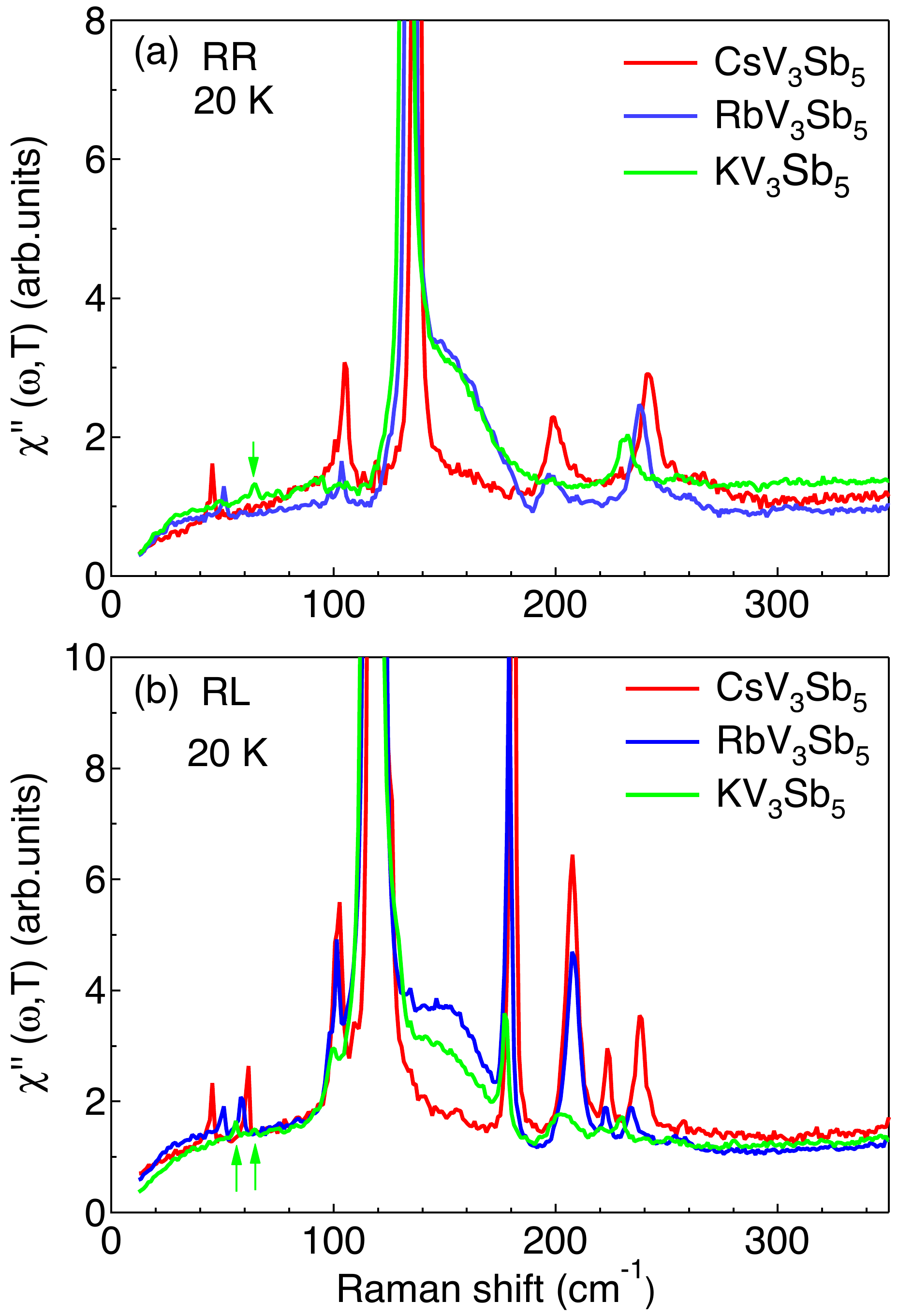}
\end{center}
\caption{\label{Fig3_AV3Sb5_compare}
(a) Comparison of the Raman spectra taken in $RR$ [$A_{1g}$+$A_{2g}$] scattering geometry for the three compounds CsV$_3$Sb$_5$, RbV$_3$Sb$_5$, KV$_3$Sb$_5$. (b) Same as (e) but in $RL$ [$E_{2g}$] scattering geometry.
The green arrows in (a) and (b) locate the weak low-energy phonons for KV$_3$Sb$_5$.
}
\end{figure}

From the group-theoretical considerations of the high-temperature structure, $\Gamma$-point phonon modes of the hexagonal AV$_3$Sb$_5$ are expressed as $\Gamma_\text{total}$ = $A_{1g}$ + 4$A_{2u}$ + $B_{1g}$ + $B_{1u}$ + 2$B_{2u}$ + 2$E_{2u}$ + $E_{2g}$ + 5$E_{1u}$+ $E_{1g}$. Raman active modes are $\Gamma_{\text{Raman}}$= $A_{1g}$ + $E_{2g}$ + $E_{1g}$.
Note that $A_{1g}$ and $E_{2g}$ phonons can be measured from the $ab$-plane measurement while the $E_{1g}$ phonon can only be accessed from the $ac$ surface.

In Figs.~\ref{Fig2_CDW_phonon}(a)-\ref{Fig2_CDW_phonon}(d), we compare the phonon modes of CsV$_3$Sb$_5$ above and below $T_S$ (94\,K) measured in four different scattering geometries.  Above $T_S$, as shown in Fig.~\ref{Fig2_CDW_phonon}(a), the sharp phonon mode at around 119\,\cm-1detected in $XX$, $XY$, and $RL$ scattering geometries corresponds to the $E_{2g}$ phonon, which is the in-plane lattice vibration of the Sb atoms in the honeycomb net. The other sharp phonon at 137\,\cm-1 detected in $XX$ and $RR$ scattering geometries corresponds to the $A_{1g}$ phonon. It is the $c$-axis lattice vibration of the Sb atoms in the honeycomb net. 
As we show the symmetry decompositions in Fig.~\ref{Fig2_CDW_phonon}(b), the $E_{2g}$ and $A_{1g}$ phonon modes are clearly separated into the $E_{2g}$ and $A_{1g}$ channels, respectively.
                                                                                               
Below $T_S$ (94\,K), new phonon modes are expected to appear in the CDW state~\cite{Devereaux2007RMP}.
While there are 26$A_{1g}$ + 33$E_{2g}$ + 35$E_{1g}$ Raman-active modes in the $2\times2\times4$ structure due to BZ folding (Appendix~\ref{GroupTheory}), only those modes modulating the ionic deviation from the high-temperature structure [Fig.~\ref{Fig6_displacement_pattern}] with a large amplitude can gain noticeable Raman intensity, thus can be detected in the Raman spectra~\cite{Klein1982PhysRevB,Holy1976PRL,NAGAOSA1982809}. 
These modes are the amplitudes modes of the CDW order parameter. The new $A_{1g}$ modes are the amplitudes modes of the CDW ground state while the new $E_{2g}$ modes are the amplitudes modes of the `failed order parameters'~(Sec.~\ref{SUBDUCTION}).
Specifically, in the case of dual order parameters, each new $A_{1g}$ mode is a doublet at low temperatures~(Sec.~\ref{Landau}).
     
In Figs.~\ref{Fig3_T_dependence} and ~\ref{Fig2_CDW_phonon}(c), we show several new phonon modes appearing in all four scattering geometries below $T_S$. 
The intensities of these new phonon modes are generally two orders of magnitude weaker than the main phonon peak at 119 and 137\,\cm-1, consistent with the weak superlattice Bragg peaks observed in the x-ray scattering measurements~\cite{Li_arxiv2021,Ortiz_arxiv2021}.
The symmetry of these new phonon modes can be clearly distinguished by decompositions according to the point group $D_{6h}$ shown in Fig.~\ref{Fig2_CDW_phonon}(d). This indicates that the three-fold rotational symmetry remains intact in the CDW phase~[Appendix~\ref{Raman_tensor_analysis}].

Specifically, three additional doublet modes are detected in the $A_{1g}$ channel, and seven additional modes are observed in the $E_{2g}$ channel below $T_S$. These new phonon peak positions are summarized in Table~\ref{phonon_modes221}. 
From Table~\ref{phonon_modes221}, we find that the experimentally observed phonon frequencies for CsV$_3$Sb$_5$ at 20\,K agree with the DFT phonon calculations based on the $2\times2\times1$ iSoD structure, suggesting the iSoD-type distortion is dominant in the CDW ground state. This is consistent with the major iSoD-type displacement patterns in the x-ray results~[Fig.~\ref{Fig6_displacement_pattern}].                         
       
It deserves to remark on the charge density gaps which are about $2\Delta\approx40$\,meV determined by STM~\cite{Jiang2021NatureMaterial}, ARPES~\cite{Comin_arxiv2021,ZhangYan_arxiv2021,ZhouXJ_arxiv2021}, and ultrafast measurements~\cite{Wang_arxiv2021}. The CDW gap-opening signatures, namely, the suppression of the low-energy spectra weight and the enhancement of the spectra weight close to $2\Delta\approx 40$\,meV, are not observed in the Raman response both in the $A_{1g}$ and $E_{2g}$ channels. The absence of CDW gap-opening signatures may be due to the multiband effects in AV$_3$Sb$_5$ system.

We note that if the inversion symmetry does not hold, the infrared-active $A_{2u}$ and $E_{1u}$ modes will become Raman active. However, no such modes are observed~[Table~\ref{phonon_modes221}]. Thus, the inversion symmetry remains intact in the CDW phase, consistent with the second harmonic generation measurements~\cite{Ortiz_arxiv2021,Yu_arxiv2021usR} and the x-ray diffraction results~\cite{Ortiz_arxiv2021}.  
                                                           
After establishing the new phonon modes in the CDW phase of CsV$_3$Sb$_5$, it is instructive to take a closer look at their sibling compounds RbV$_3$Sb$_5$ and KV$_3$Sb$_5$. In Fig.~\ref{Fig3_AV3Sb5_compare}(a) and \ref{Fig3_AV3Sb5_compare}(b), we compare the Raman response in $RR$ and $RL$ scattering geometries for the three compounds. In general, the spectral features for all three compounds are similar. They show a similar number of new phonon modes at similar positions at 20\,K. The summary of the new phonon modes for the three compounds at 20\,K is presented in Table~\ref{phonon_modes221}.

\begin{table}[t]
\caption{\label{phonon_modes221} The phonon frequencies calculated by DFT for the $2\times2\times1$ SoD and iSoD structures for CsV$_3$Sb$_5$ and the experimentally observed phonon frequencies at the Brillouin zone center for the three compounds AV$_3$Sb$_5$ (A=Cs, Rb, K) at 20\,K. All the units are in \cm-1.}
\begin{ruledtabular}
\begin{tabular}{ccccccc}
&CsV$_3$Sb$_5$&CsV$_3$Sb$_5$&CsV$_3$Sb$_5$&RbV$_3$Sb$_5$&KV$_3$Sb$_5$\\
Sym.& SoD& iSoD&&&\\
&(DFT)&(DFT)&(Exp)&(Exp)&(Exp)\\
\hline
&227&242&241,262&238,260&231,254\\
&200&203&197,202&196,201&\\  
$A_{1g}$&143&142&&&\\  
&124&132&137&136&133\\  
&93&108&99,105&98,104&92,95\\  
&&&45&51&64\\  
\hline
&247&234&238&234&230\\
&219&221&223&223&220\\  
&172&212&208&208&202\\  
&169&175&181&179&177\\  
$E_{2g}$&130&129&&&\\  
&120&125&119&119&117\\  
&55&99&102&101&100\\  
&50&58&61&59&56\\  
&&&45&51&64\\  
\end{tabular}
\end{ruledtabular}
\end{table}

\begin{table}[b]
\caption{\label{phonon_modes_Compare} Comparison of phonon scattering rate for the main $A_{1g}$ and $E_{2g}$ phonons at 20\,K for the three compounds  AV$_3$Sb$_5$ (A=Cs, Rb, K). The phonon scattering rate is obtained as the inverse of the HWHM. All the HWHM data have been corrected for the instrumental resolution. The unit of scattering rate is in GHz.}
\begin{ruledtabular}
\begin{tabular}{ccc}
sample	&$A_{1g}$&$E_{2g}$\\
\hline
CsV$_3$Sb$_5$	&5.4&10.2\\
RbV$_3$Sb$_5$	&6.9&13.8\\   
KV$_3$Sb$_5$		&26.7&28.5\\  
\end{tabular}
\end{ruledtabular}
\end{table}

\begin{figure}[!t] 
\begin{center}
\includegraphics[width=\columnwidth]{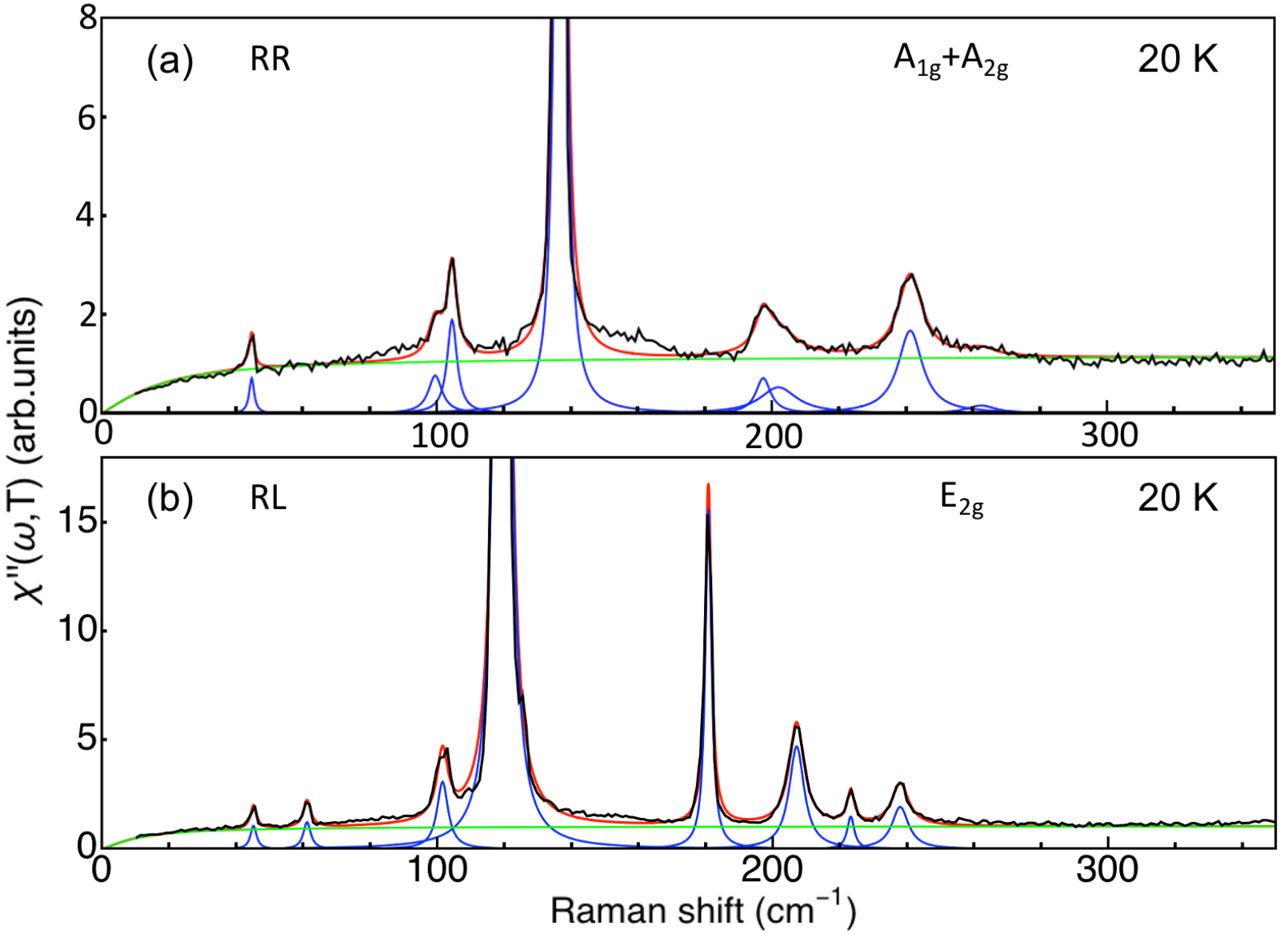}
\end{center}
\caption{\label{RR_RL_fitting_20K} 
An example of fitting for the Raman response in $RR$ (a) and $RL$ (b) scattering geometries at 20\,K for CsV$_3$Sb$_5$. The red, blue, and green lines represent the total fitted response, the individual Lorentzian components, and a smooth background, respectively.}
\end{figure}

According to the group-theoretical analysis, for the $2\times2\times1$ SoD or iSoD phase, all the Raman-active phonon modes are related to the V atoms in the kagome net and Sb atoms in the honeycomb net. The lattice vibration for the alkali-metal atoms (Cs, K, Rb) are forbidden. However, they become Raman-active in the $2\times2\times4$ structure~[Appendix~\ref{GroupTheory}].
Based on Table~\ref{phonon_modes221}, all the phonon frequencies decrease when switching from Cs to K, except  for the low-energy modes showing both $A_{1g}$ and $E_{2g}$ symmetries, e.g.~45\,\cm-1 mode in CsV$_3$Sb$_5$. 
The distinct behavior for this phonon with $A_{1g}$ and $E_{2g}$ symmetry indicates it does not come from the kagome net nor from the honeycomb net. 
Based on the fact that (1) Cs is the heaviest atom in CsV$_3$Sb$_5$, it should have the lowest vibration frequency; (2) the atomic mass deceases when switching from Cs to K in AV$_3$Sb$_5$, the vibration frequency is expected to increase, we conclude that the low-energy phonon modes at 45\,\cm-1 in CsV$_3$Sb$_5$, 51\,\cm-1 in RbV$_3$Sb$_5$, and 64\,\cm-1 in KV$_3$Sb$_5$ must be the alkali-atom-related lattice vibration modes [Appendix~\ref{GroupTheory}].
Furthermore, these modes appear for all four scattering geometries, thus they do not obey the $D_{6h}$ selection rules~[Table~\ref{SymmetryAnalysis} and \ref{decompositionD6h}], suggesting they are not bulk phonons.
Moreover, the amplitude of the mode at 45\,\cm-1 in CsV$_3$Sb$_5$ shows two-fold rotational symmetry in the
pump-probe Kerr rotation measurement at 20\,K~\cite{Wu_Kerr_arxiv2021}, indicating a $C_3$-symmetry-broken phase.
This ``nematiclike" $C_3$-symmetry-broken phase would be characterized by the failed order parameter with $\Gamma^+_5$ ($E_{2g}$) symmetry that is stabilized on the surface~[Section~\ref{SUBDUCTION}]. Indeed, the alkali-metal surface reconstruction, namely, the half alkali-metal surface, is most commonly observed in STM experiments~\cite{Liang_arxiv2021}.
                                                                                                                                                                                                                                    
However, we notice several quantitative differences for the three compounds AV$_3$Sb$_5$ (A=Cs, Rb, K). 
First, the intensity of the new phonon modes are the strongest in CsV$_3$Sb$_5$ among the three compounds; it becomes weaker in RbV$_3$Sb$_5$ and becomes much weaker in KV$_3$Sb$_5$. 
We note that the amplitude of specific jump at $T_S$ for KV$_3$Sb$_5$ is also the smallest among the three systems~\cite{Ortiz2019PhysRevMaterials,Ortiz2020PhysRevLett,Ortiz2021PhysRevMaterials,Yin_2021_CPL}.
Second, the scattering rate for the main phonon modes is the smallest for CsV$_3$Sb$_5$ and the largest for KV$_3$Sb$_5$~[Table~\ref{phonon_modes_Compare}]. This might explain why the new phonon intensity in KV$_3$Sb$_5$ is weaker, as there might be more disorders or stacking faults in KV$_3$Sb$_5$ than the other two systems. 
Third, there is a broad peak centered at around 150\,\cm-1 for RbV$_3$Sb$_5$ and KV$_3$Sb$_5$ in both $RR$ and $RL$ scattering geometries~\cite{Li_arxiv2021,Wulferding_arxiv2021}, but not noticeable for CsV$_3$Sb$_5$.
                                                                                    
\subsubsection{Temperature dependence of main modes}\label{Temperature_dependence}                                                                                                
After establishing the phonon modes in the three AV$_3$Sb$_5$ systems, we switch to study the $T$ dependence of these modes.
                
In Fig.~\ref{Fig3_T_dependence}, we present the $T$ dependence of the phonon modes in both $RR$ and $RL$ scattering geometries for CsV$_3$Sb$_5$. The main $E_{2g}$ phonon at 119\,\cm-1 and main $A_{1g}$ phonon at 137\,\cm-1 persist across $T_S$. In contrast, all the new $E_{2g}$ modes abruptly appear below $T_S$. 
Most of the new $A_{1g}$ modes first appear as relatively broad features which then sharpen upon cooling.

 \begin{figure}[!b] 
\begin{center}
\includegraphics[width=\columnwidth]{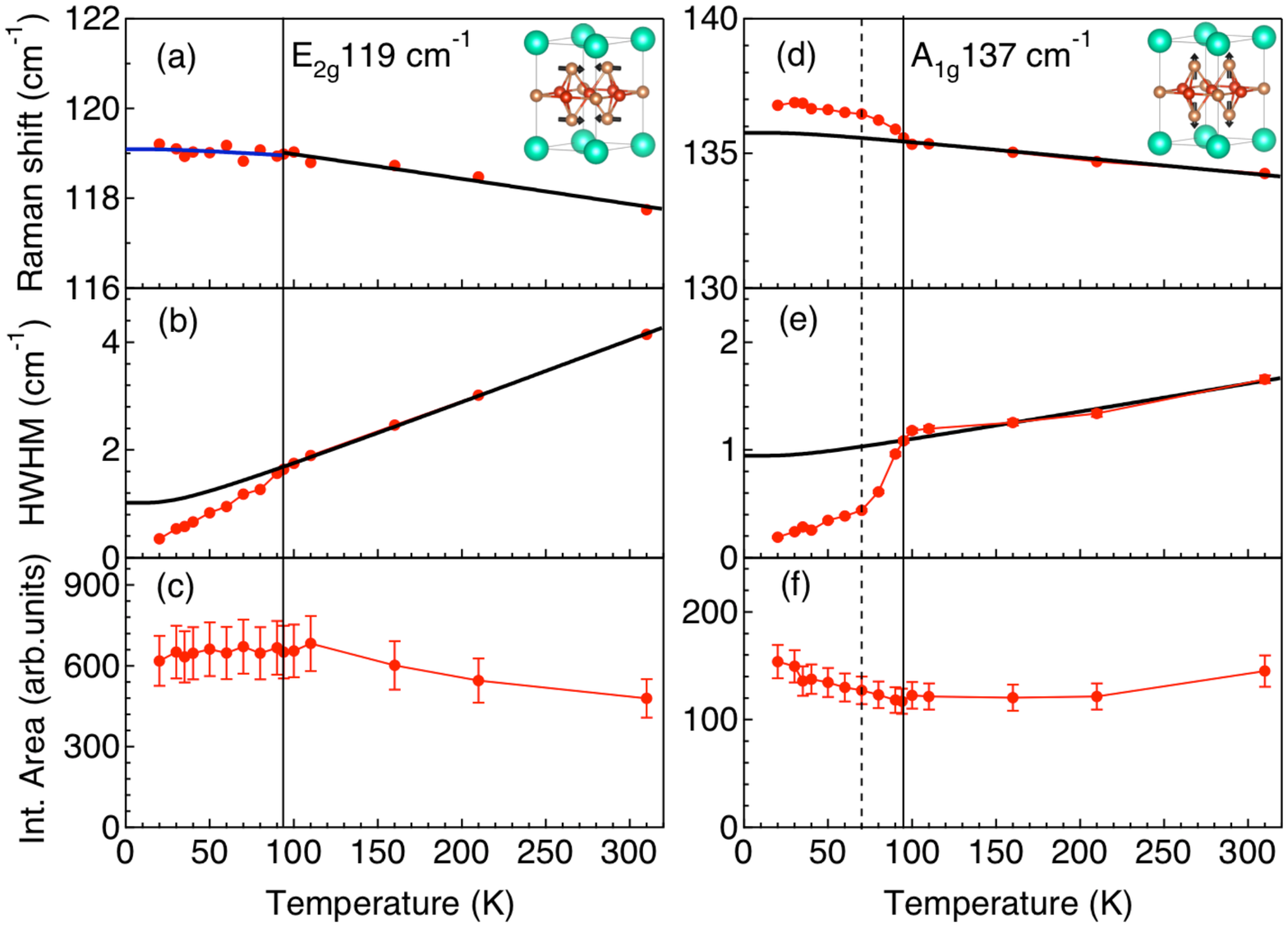}
\end{center}
\caption{\label{Main_mode1} 
$T$-dependence of the peak position (a), HWHM (half-width at half-maximum) (b) and integrated intensity (c) for the main $E_{2g}$ phonon at 119\,\cm-1 for CsV$_3$Sb$_5$.
(d)-(f) Same as (a)-(c) but for the main $A_{1g}$ phonon at 137\,\cm-1. The error bars represent one standard deviation.
The solid black and blue lines represent the fitting of the phononic self-energy $T$ dependence above $T_S$ and below $T_S$ by anharmonic decay model, respectively [Eqs.~(\ref{eq_omega1}) and (\ref{eq_gamma1})]. The solid vertical lines represent $T_S$ while the dashed black lines represent $T^*$. The insets of (a) and (d) show the lattice vibration patterns for the $E_{2g}$ phonon and $A_{1g}$ phonon, respectively.}
\end{figure}

We fit the Raman spectra shown in Fig.~\ref{Fig3_T_dependence} by multi-Lorentzian peaks on a smooth background. Examples of the fits for the Raman response in $RR$ and $RL$ scattering geometries at 20\,K are presented in Fig.~\ref{RR_RL_fitting_20K}. 

In Fig.~\ref{Main_mode1}, we show the $T$ dependence of the phonon frequencies, HWHM, and the integrated intensity for the main $E_{2g}$ phonon at 119\,\cm-1 and main $A_{1g}$ mode at 137\,\cm-1 in CsV$_3$Sb$_5$.                                              
For the main $E_{2g}$ phonon, the frequency increases upon cooling and shows little changes across $T_S$. 
It can be well described by the anharmonic phonon decay model~[Appendix~\ref{Anharmonic_decay_model}].
The HWHM decreases upon cooling and decreases faster below $T_S$. The integrated intensity increases slowly upon cooling and stays almost a constant below $T_S$.
                                                            
The main $A_{1g}$ phonon hardens upon cooling and shows a large additional hardening below $T_S$, suggesting the main $A_{1g}$ phonon couples to the $A_{1g}$-symmetry CDW order parameter below $T_S$.         
The main $A_{1g}$ phonon narrows upon cooling and narrows much faster below $T_S$. 
The HWHM of this mode starts to decrease slower at a lower temperature $T^*=70$\,K. The $T$-dependence of HWHM for this mode strongly deviates from the anharmonic phonon decay model~[Appendix~\ref{Anharmonic_decay_model}], suggesting phononic self-energy effect due to the coupling to the $A_{1g}$-symmetry CDW order parameter below $T_S$.
The integrated intensity for the main $A_{1g}$ phonon decreases a bit upon cooling and increases slightly below $T_S$, which is an indication of finite electron-phonon coupling in this system~\cite{Mai2019PRB}.

\begin{figure*}[!t] 
\begin{center}
\includegraphics[width=1.7\columnwidth]{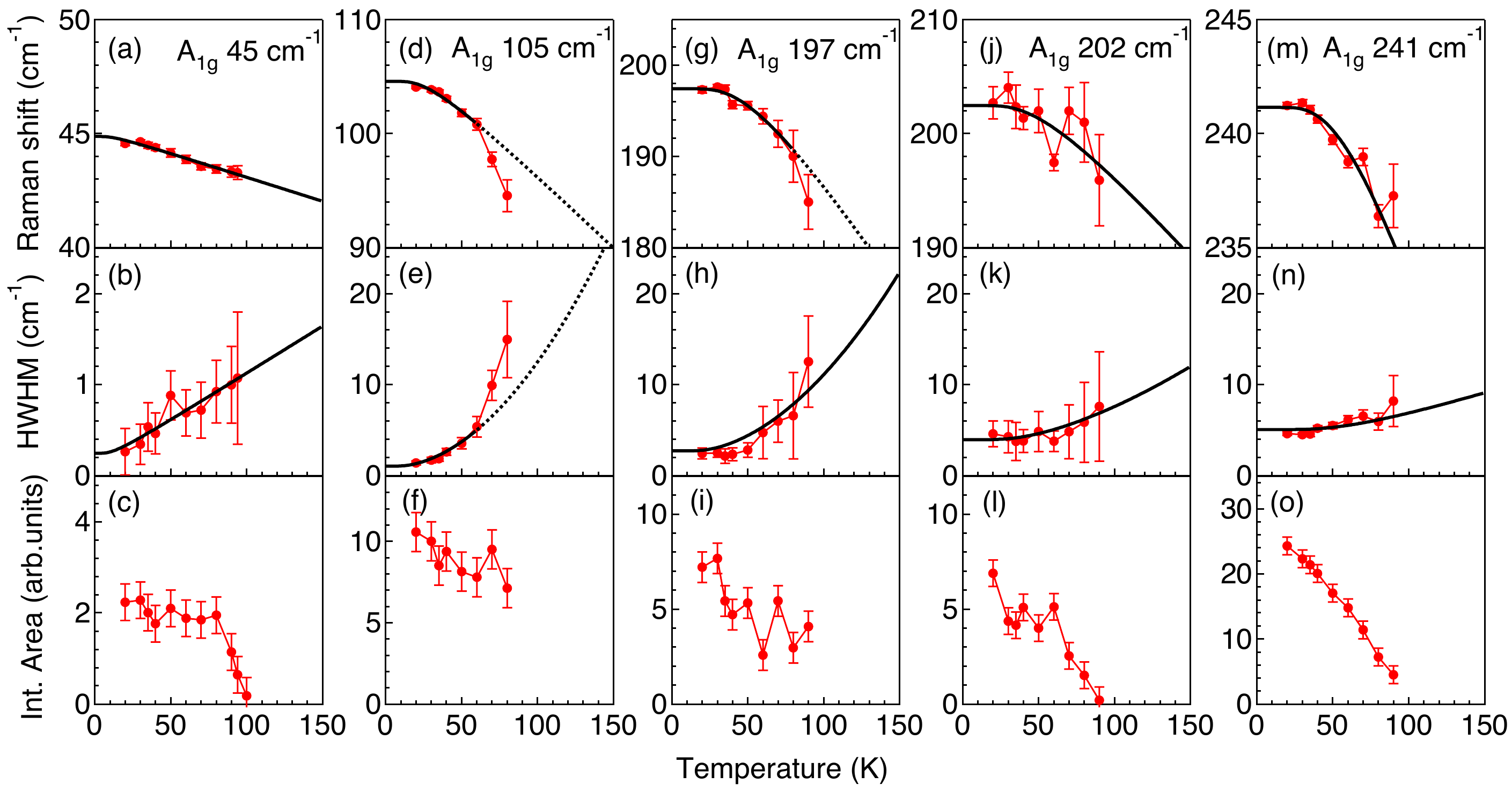}
\end{center}
\caption{\label{A1g_new_modes_with_correction} 
$T$-dependence of the peak position, HWHM, and integrated intensity for the new $A_{1g}$ phonon mode below $T_S$ for CsV$_3$Sb$_5$.
(a)-(c) For the mode at 45\,\cm-1.
(d)-(f) For the mode at 105\,\cm-1.
(g)-(i) For the mode at 197\,\cm-1.
(j)-(l) For the the mode at 202\,\cm-1.
(m)-(o) For the mode at 241\,\cm-1.
The error bars represent one standard deviation.
For fitting of the phononic self-energy $T$-dependence for the mode at 45 and 241\,\cm-1, Eqs.~(\ref{eq_omega1}) and Eq.~(\ref{eq_gamma1}) are used. For the modes at 105, 197, and 202\,\cm-1,
Eqs.~(\ref{eq_omega2}) and (\ref{eq_gamma2}) are used.}
\end{figure*}

\subsubsection{Temperature dependence of the amplitude modes}\label{New_modes}

In this section, we study the $T$ dependence of the new $A_{1g}$ and $E_{2g}$ modes. They are both the amplitude modes of the CDW order parameter.

We first discuss the $T$-dependence of the new $A_{1g}$ amplitude modes. Since each new $A_{1g}$ mode is a doublet~[Section~\ref{Landau}], we fit each new $A_{1g}$ mode with double-Lorentzian functions except for the 45\,\cm-1 mode~[Fig.~\ref{RR_RL_fitting_20K}(a)].                                                        

In Fig.~\ref{A1g_new_modes_with_correction}, we present the selected $T$ dependence of peak frequency, HWHM, and integrated intensity for the new  $A_{1g}$ phonon modes. For the modes at 45, 202, and 241\,\cm-1, the $T$ dependence of these peak frequency and HWHM follow the anharmonic phonon decay model well below $T_S$~[Appendix~\ref{Anharmonic_decay_model}]. 
To the contrary, the $T$-dependence of the phonon peak frequency and HWHM for the mode at 105\,\cm-1 deviates from the anharmonic phonon decay model in the range $60\sim80$\,K~[Fig.~\ref{A1g_new_modes_with_correction}(d) and (e)]. Same is found for the mode at 197\,\cm-1 which shows a deviation at around $T^*$=$70$\,K, as shown in Fig.~\ref{A1g_new_modes_with_correction}(g).  
These anomalies are consistent with the theoretical $T$-dependence of the amplitude modes frequencies at around 80\,K~[Fig.~\ref{Order_parameter}(d)], which is close to $T^*$ that is set to be 70\,K in the free-energy model~[Eq.~\ref{free_energy}]. 
These anomalies at around $T^*$ results from the interplay between the primary-like and secondary-like order parameters.
                                                                                                                                                                                                                                                                 
The integrated intensity for the new $A_{1g}$ mode at 241\,\cm-1 increases monotonically below $T_S$~[Fig.~\ref{A1g_new_modes_with_correction}(o)], indicating a dominant order parameter emerging below $T_S$. 
To the contrary, the new $A_{1g}$ 
mode at 105\,\cm-1 first appears as a weak and broad feature below $T_S$, then becomes noticeable at around $T^*=70$\,K and gradually gain intensity below $T^*$~[Fig.~\ref{Fig3_T_dependence}(b)].  
The integrated intensity for this mode shows a saturation-like behavior below $T^*$~[Fig.~\ref{A1g_new_modes_with_correction}(f)]. Similar behaviors are also found for the modes at 197\,\cm-1 and 202\,\cm-1~[Fig.~\ref{A1g_new_modes_with_correction}(i) and (l)].
These results suggest that a secondary order parameter comes into play below $T^*$.

The above two different behaviors for the $A_{1g}$ modes' integrated intensity --the monotonical increase and the saturation-like behavior-- can be captured by the Landau free-energy model~[Eq.~\ref{free_energy}] incorporating the interplay between the primary-like ($u_0$) and the secondary-like ($l_0$) order parameters via trilinear coupling $\gamma u_0^2 l_0$.
The plateau-like behavior for the $A_{1g}$ mode at 105\,\cm-1 below $T^*$ can be modeled by the $T$-dependence of $(u_0+l_0)^2$ in the case of $\gamma>0$ where $u_0$ and $l_0$ have different signs. In this case, $u_0$ and $l_0$ interplay with each other destructively, thus $(u_0+l_0)^2$ shows a plateau-like behavior below $T^*$~[Fig.~\ref{Order_parameter}(e)].
The monotonic increase of the integrated intensity for the higher-energy $A_{1g}$ modes at 241\,\cm-1 can be qualitatively
described by the $T$-dependence of the $(u_0+l_0)^2$ in the case of $\gamma<0$ where $u_0$ and $l_0$ have the same signs. In this case, $u_0$ and $l_0$ interplay with each other constructively, thus $(u_0+l_0)^2$ increases monotonically below $T_S$~[Fig.~\ref{Order_parameter}(f)].

The above two different behaviors for the integrated intensities originate from the multiband nature of AV$_3$Sb$_5$. Recent angle-resolved photoemission measurements indeed revealed that multiple V $3d$ bands cross the Fermi level~\cite{Comin_arxiv2021,ZhangYan_arxiv2021,ZhouXJ_arxiv2021,Wang_arxiv2021}. 
The Landau free energy parameters, especially the trilinear coupling constant $\gamma$, vary for different bands. They determine whether the primary-like $U_1$ and secondary-like $L_{2}^{-}$ order parameters interplay constructively or destructively, thus determining the shape of the $T$-dependence of the integrated intensity below $T_S$.

We note that the appearance of the $A_{1g}$ mode at 105\,\cm-1 below $T^*$ was also reported in the ultrafast coherent phonon spectroscopy measurements ~\cite{Ratcliff_arxiv2021,Wang_arxiv2021,Wu_Kerr_arxiv2021}. 
The authors of these works~\cite{Ratcliff_arxiv2021,Wang_arxiv2021,Wu_Kerr_arxiv2021} linked the temperature $T^*$ to the emergence of a uniaxial charge modulation observed by STM~\cite{Liang_arxiv2021,Zhao_arxiv2021,Chen_arxiv2021}, which breaks both $C_6$/$C_3$ rotational symmetry.
However, the presented here x-ray and Raman data do not support this scenario. First, both x-ray and Raman results indicate that $C_3$ rotational symmetry is preserved in the CDW ground state. Secondly, the Raman data shown in Fig.~\ref{Fig3_T_dependence} do not show any additional sets of new phonon modes below $T^*$. Third, the refined low-temperature structure shown in Fig.~\ref{Fig6_displacement_pattern} does
not contain an interlayer $\pi$ phase shift, ruling out the
bulk $D_{2h}$ $2\times2\times2$ CDW order~\cite{miao2021geometry}.
Thus, the presented here results suggest that the uniaxial charge modulation below $T^*$ is not a bulk effect.

\begin{figure*}[!t] 
\begin{center}
\includegraphics[width=2\columnwidth]{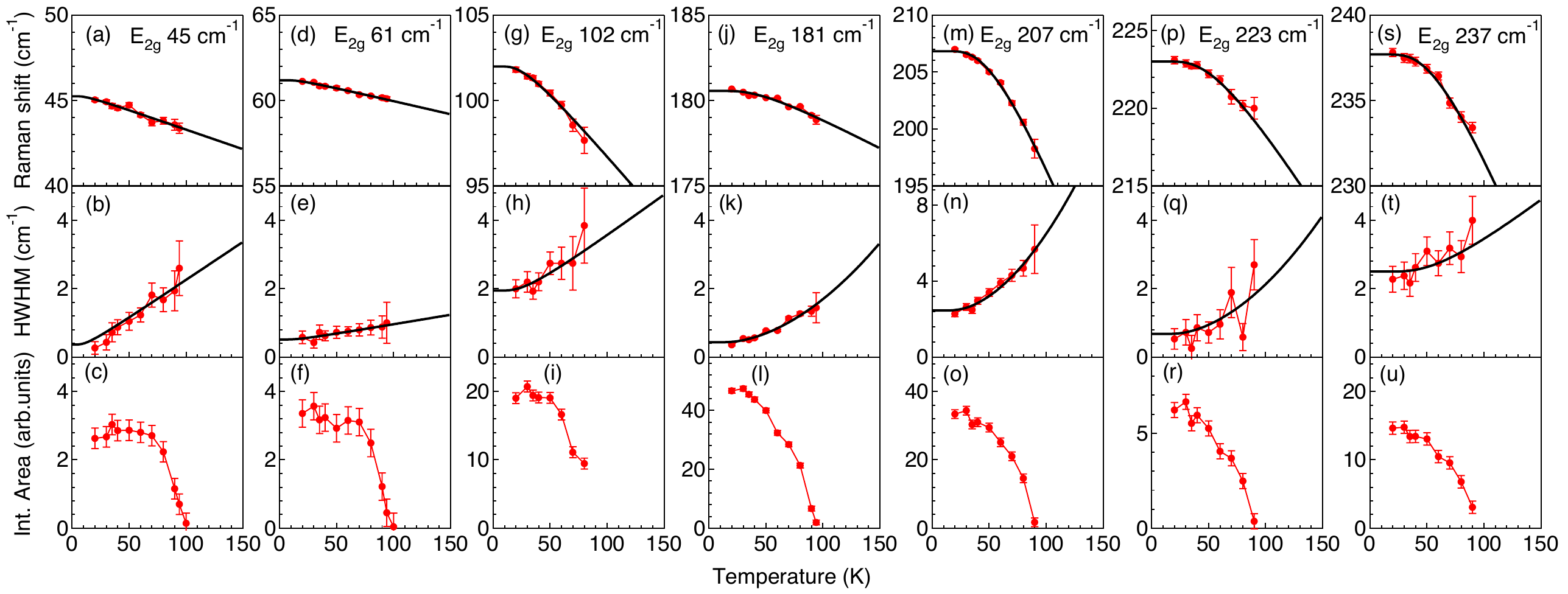}
\end{center}
\caption{\label{E2g_new_modes_with_correction} 
$T$ dependence of the peak position, HWHM, and integrated intensity for the new $E_{2g}$ phonon modes below $T_S$ for CsV$_3$Sb$_5$. 
(a)-(c) For the mode at 45\,\cm-1. 
(d)-(f) For the mode at 61\,\cm-1.
(g)-(i) For the mode at 102\,\cm-1.
(j)-(l) For the mode at 181\,\cm-1.
(m)-(o) For the mode at 207\,\cm-1.
(p)-(r) For the mode at 223\,\cm-1.
(s)-(u) For the mode at 237\,\cm-1.
The error bars represent one standard deviation.
For fitting of the phononic self-energy $T$-dependence for the modes at 45, 61, and 102\,\cm-1, Eq.~(\ref{eq_omega1}) and Eq.~(\ref{eq_gamma1}) are used. For the modes at 181, 207, and 223\,\cm-1, Eq.~(\ref{eq_omega1}) and (\ref{eq_gamma1}) are used.
}
\end{figure*}    
         
Finally, we switch to discuss the $T$ dependence of the $E_{2g}$ modes, which are accessed spectroscopically by exciting the amplitude modes of the failed order parameters.
In Fig.~\ref{E2g_new_modes_with_correction}, we present the $T$ dependencies of the new $E_{2g}$ phonon modes. In general, the $T$ dependencies of peak frequency and HWHM are smooth, showing no anomalies below $T_S$. They follow the anharmonic phonon decay model well below $T_S$~[Appendix~\ref{Anharmonic_decay_model}].
The abrupt appearance of the new $E_{2g}$ phonon modes below $T_S$ can be clearly seen from the $T$ dependence of the integrated intensities. They onset at $T_S$ and increase monotonically below $T_S$, except for the low-energy $E_{2g}$ mode at 61\,\cm-1. 
The integrated intensity for this mode shows a saturationlike behavior below $T^*$=70\,K. 
These two distinct behaviors are similar to the temperature dependence of the $A_{1g}$ amplitude modes shown in Fig.~\ref{A1g_new_modes_with_correction}, consistent with the primary-like and secondary-like order parameters.

\section{Conclusion}\label{Conclusion}
                        
In summary, we study the primary and secondary order parameters associated with the CDW transition in the kagome metal AV$_3$Sb$_5$ system by polarization-resolved electronic Raman spectroscopy and DFT calculations.
                                                                                                                        
Previous x-ray diffraction data at 15\,K established that the CDW order in CsV$_3$Sb$_5$ is a $2\times2\times4$ structure with space group $P\bar{3}$: one layer of iSoD structure, and three consecutive layers of SoD structure with zero-phase-shift between neighboring layers. 
This $2\times2\times4$ structure can be further refined to space group $P6/mmm$ if we disregard the tiny distortions.  
The vanadium atoms show major displacements in the iSoD layer while they show minor displacements in the SoD layer. By quantitatively analyzing the vanadium lattice distortions, we identify that $U_1$ lattice distortion is the primary-like order parameter while $M^+_1$ and $L^-_2$ distortions are secondary-like order parameters.
This is also confirmed by the calculation of bare static susceptibility $\chi'_0(q)$ that shows a broad peak at around $q_z=0.25$ along the hexagonal Brillouin zone face central line ($U$ line).                                                     
                                                                                                                                                
The primary and secondary order parameters are revealed by Raman scattering.
We identify several $A_{1g}$ and $E_{2g}$ phonon modes related to V and Sb atoms as well as alkali-metal atoms emerging in the CDW state. 
The symmetry decompositions analysis for these V-and-Sb-atoms-related modes indicates the $C_3$ symmetry is preserved in the CDW phase. To the contrary, the low-energy alkali-metal-atom-related lattice vibration modes do not obey the $D_{6h}$ point group selection rules.             
These alkali phonon modes indicates a ``nematiclike" $C_3$-symmetry-broken phase characterized by the ``failed order parameter" with $\Gamma^+_5$ ($E_{2g}$) symmetry that is stabilized on the surface.
    
The observed number of Raman-active phonon modes in the CDW state indicates that the inversion symmetry remains intact.
By comparing the DFT phonon calculations and the observed new phonon frequencies, we find that the $2\times2\times4$ structure hosts a dominant iSoD-type lattice distortion, consistent with the x-ray refinement results.
                                                                                                                                                                                                                                                                                                                                              
The detailed temperature evolution of these new modes' peak frequency, HWHM, and integrated intensity support two successive phase transitions in CsV$_3$Sb$_5$: the first one associated with the primary-like order parameter appearing at $T_S=94$\,K and the second isostructural one appearing at $T^*=70$\,K. Moreover, the $T$-dependence of the integrated intensity for these modes show two type of behavior below $T_S$: a plateau-like behavior below $T^*$ and monotonically increase below $T_S$.
These two behaviors are captured by a Landau free-energy model incorporating the interplay between the primary-like and the secondary-like order parameters via trilinear coupling. Especially, the sign of the trilinear term determines whether the primary-like and secondary-like order parameters cooperate or compete with each other, thus determining the shape of the $T$ dependence of the Raman data below $T_S$.

These results establish a solid foundation to study the interplay between the primary and secondary CDW order parameters in the kagome metal system. They guide to identify the primary and secondary order parameters as well as their interplay, when these CDW order parameters are tuned by carrier doping~\cite{Oey_2021arXiv,Liu_Ti_2021arXiv,Song2021PhysRevLett}, external pressure~\cite{Yu2021,ChenPhysRevLett.126.247001}, or strain~\cite{Qian_arxiv2021}.

\textit{Note added in proof}. Recently, G. Liu $et~al$. have posted Raman data for CsV$_3$Sb$_5$ which is consistent with the data reported here~\cite{Liu_2021arXivRaman}. In contrast, polarization-insensitive broad feature at about 150~\cm-1 with anomalous temperature dependence reported and studied by H. Li $et~al$.~\cite{Li_arxiv2021}, or by D. Wulferding $et~al$.~\cite{Wulferding_arxiv2021} is inconsistent with presented here data, thus must be of an external origin. 
                                        
                     
\begin{acknowledgments}
The spectroscopic work conducted at Rutgers 
(S.-F.W. and G.B.) was supported by NSF Grants No.~DMR-1709161 and No.~DMR-2105001. 
The sample growth and characterization work
 conducted at UC Santa Barbara (B.R.O. and S.D.W.) was supported by the UC Santa Barbara NSF Quantum Foundry funded via the Q-AMASE-i program under Award No.~DMR-1906325. 
B.R.O. also acknowledges support from the California NanoSystems Institute through the Elings fellowship program.
The DFT phonon and bare susceptibility calculations work conducted at Weizmann Institute of Science (H.X.T. and B.H.Y.) was supported by the European Research Council (ERC Consolidator Grant ``NonlinearTopo'', No.~815869), the ISF-Quantum Science and Technology (No.~1251/19).
The theoretical work conducted at the University of Minnesota (T.B.) was supported by the NSF~CAREER Grant No.~DMR-2046020.
The work at NICPB was supported by the European Research Council (ERC) under the European Union’s Horizon 2020 research and innovation programme grant agreement No.~885413.

\end{acknowledgments}     
     
     
\appendix

\section{Laser heating determination} \label{laser_heating_determination}                                                                                                                             
The laser heating rate, a measure of the temperature increase per unit laser power (K/mW) in the focused laser spot, in the Raman experiments was determined by monitoring 
the appearance of new phonon modes induced by the CDW order during the cooling process with a constant laser power 10\,mW.
         
At the cryostat temperature 90\,K, we barely detect any new phonon modes, indicating the laser spot temperature is above $T_{S}$=94\,K.
When cooling the sample to 85\,K, we start to detect several weak new phonon signals both in the $RR$ and $RL$ scattering geometries, indicating the laser spot temperature is slightly below 94\,K. When cooling the sample to 80\,K, the intensity of these new modes develop significantly, indicating the laser spot temperature is well below 94\,K. 
Thus, the heating coefficient can be determined via: $85+10*k \approx 94$. In this way, we have deduced the heating coefficient: $k \approx 0.9\pm0.1$\,K/mW.

\section{Removal of polarization leakage}\label{leakage}

In this appendix, we provide the detailed procedure to remove the polarization leakage signal
from optical elements in our data analysis. 

In our polarization optics setup, we used a Glan-Taylor polarizing prism (Melles Griot) with an extinction ratio better than $10^{-5}$ to clean the laser excitation beam and a broad-band 50\,mm polarizing cube (Karl Lambrecht Corporation) with an extinction ratio better than 1:500 to analyze the scattered light.
For the linearly polarized $XX$ and $XY$ scatting geometries, the leakage intensity ratio is negligibly small (less than 0.2\,\%), thus the leakage intensity from the linearly polarized scattering geometries is not considered.

For measurements with circularly polarized light, we employed a Berek compensator (New Focus) to convert the incoming linearly polarized light into circularly polarized light for excitation. We used a broad-band 50-mm-diameter quarter wave retarder (Melles Griot) with a retardance tolerance $\lambda/50$ before the polarizing cube to convert the out-coming circularly polarized light into linearly polarized light for the analyzer. 
The leakage of circular polarized light is due to the limitations of the broadband quarter wave plate and
alignment of the Berek compensator.

The amount of phonon leakage intensity is determined based on the
$A_{1g}$  and $E_{2g}$ bulk phonons of CsV$_3$Sb$_5$ at the same temperature in $RR$ and $RL$ scattering geometries. To remove the polarization leakage intensity, we subtract intensity from the orthogonal polarization
geometry, i.e., $\chi''_{RR}(\omega,T) =\overline{\chi''_{RR}}(\omega,T) - \alpha \cdot \overline{\chi''_{RL}}(\omega,T)$, where $\overline{\chi''_{RR}}(\omega,T)$ and $\overline{\chi''_{RL}}(\omega,T)$ are the raw data taken in $RR$ and $RL$ polarization scattering
geometries at temperature $T$, respectively, and $\alpha$ is a small number representing the leakage ratio. It is expected that the same ratio $\alpha$ also applies to $RL$ polarization scattering geometry as well:
$\chi''_{RL}(\omega,T) =\overline{\chi''_{RL}}(\omega,T) - \alpha \cdot \overline{\chi''_{RR}}(\omega,T)$. 
                                                                                                                                                                                                                                      
\begin{figure}[!t] 
\begin{center}
\includegraphics[width=\columnwidth]{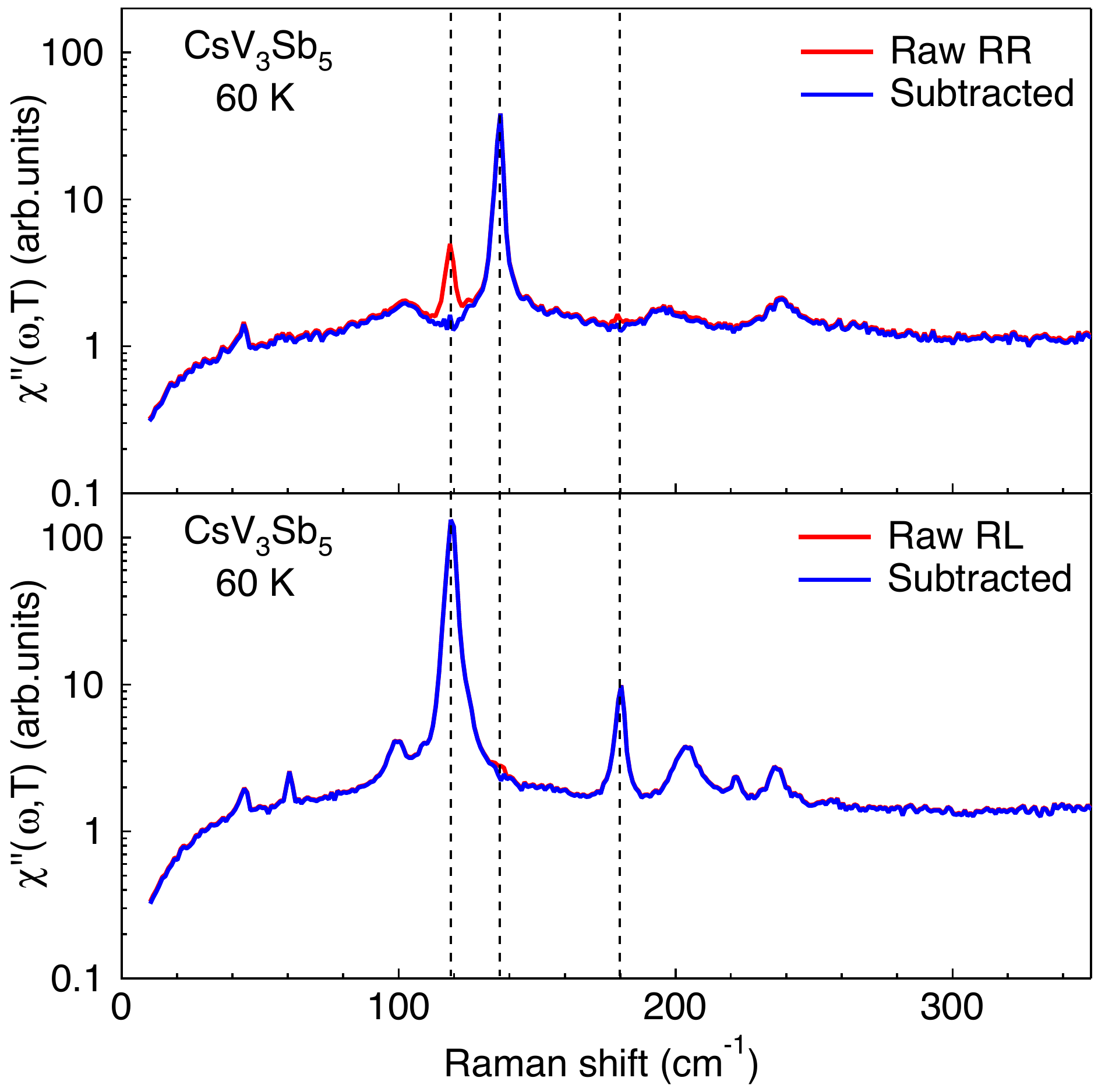}
\end{center}
\caption{\label{CsV3Sb5_remove_leakage} Comparison of raw data and polarization leakage removed spectra, taken in $RR$ (top panel) and $RL$ (bottom panel) polarization scattering geometry from the $ab$ surface of CsV$_3$Sb$_5$ at 60\,K with 647\,nm  laser excitation. }
\end{figure}

In Fig.~\ref{CsV3Sb5_remove_leakage}, we show the Raman spectra of the unprocessed raw data and
polarization-leakage-removed spectra taken at 60\,K from the
$ab$ surface of CsV$_3$Sb$_5$ crystals in $RR$ and $RL$ scattering geometries,
respectively. The leakage intensity of the bulk $E_{2g}$
phonons at 119 and 181\,\cm-1 in the raw data taken with $RR$ scattering geometries can be fully removed with a leakage ratio $\alpha$ close to $0.02$.

\section{Raman tensor analysis}\label{Raman_tensor_analysis}  
                   
The Raman tensor $R_{\mu}$ is a $2\times 2$ matrix for an irreducible representation ($\mu$) of a point group. It can be  used to estimate the phonon intensity in a Raman experiment.
With incident and scattering light polarization unit vector respectively defined as $\hat{e}_i$ and $\hat{e}_s$, the phononic Raman response is described as~\cite{Loudon_book}:
\begin{equation}
\label{eq_chi}
\chi''_{\hat{e}_i \hat{e}_s}\sim \sum_\mu|\hat{e}_i R_\mu \hat{e}_s |^2.
\end{equation}

The Raman tensors $R_{\mu}$ ($\mu=A_{1g}, A_{2g}, E_{1g}, E_{2g}$) for the irreducible representations ($\mu$) of point group $D_{6h}$ have the following forms:
\begin{displaymath}   
\left(\begin{array}{ccc}
a & 0  &0\\
0 &a &0\\
0 & 0 &b
\end{array}\right),\\
\left(\begin{array}{ccc}
0 & c  &0\\
-c &0 &0\\
0 & 0 &0
\end{array}\right),
\left(\begin{array}{ccc}
0 & 0  &0\\
0 &0 &d\\
0 & e &0
\end{array}\right)
\left(\begin{array}{ccc}
0 & 0  &-d\\
0 &0 &0\\
-e & 0 &0
\end{array}\right),\\
\end{displaymath}  
\begin{displaymath}   
\left(\begin{array}{ccc}
0 & f  &0\\
f &0 &0\\
0 & 0 &0
\end{array}\right)
\left(\begin{array}{ccc}
f & 0  &0\\
0 &-f &0\\
0 & 0 &0
\end{array}\right).
\end{displaymath}

We choose $\hat{e}_i$ and $\hat{e}_s$ to be $X$, $Y$, $R$, and $L$, where $X=(1~0~0)$, $Y=(0~1~0)$, $R=1/\sqrt{2} ~(1~i~0)$, and $L=1/\sqrt{2}~(1~-i~0)$. 

Based on Eq.~(\ref{eq_chi}), we obtain: 
\begin{equation}
\label{scattering_geometryD6h}
\begin{split}     
\chi''^{D_{6h}}_{XX}&=a^2+f^2,\\
\chi''^{D_{6h}}_{XY}&=c^2+f^2,\\
\chi''^{D_{6h}}_{RR}&=a^2+c^2,\\
\chi''^{D_{6h}}_{RL}&=2f^2.\\
\end{split}
\end{equation}
Thus, the Raman selection rules for the $D_{6h}$ point group indicate that the $XX$, $XY$, $RR$, and $RL$ polarization geometries probe the $A_{1g} + E_{2g}$, $A_{2g} + E_{2g}$, $A_{1g} + A_{2g}$, 2$E_{2g}$ symmetry excitations, respectively~[Table~\ref{SymmetryAnalysis}]. 
                                                                                                                   
The sum rule that 
$\chi''^{D_{6h}}_{XX}$ + $\chi''^{D_{6h}}_{XY}$ = $\chi''^{D_{6h}}_{RR}$ + $\chi''^{D_{6h}}_{RL}$ = $a^2+c^2+2f^2$
set a constraint for the Raman response in different scattering geometries, 
thus providing a unique way to check the data consistency.
                                                                                                                                                 
From Eq.~(\ref{scattering_geometryD6h}), we can calculate the square of the Raman tensor element:
\begin{equation}
\label{decompossiitonD6hEQ}
\begin{split}     
a^2&=\chi''^{D_{6h}}_{XX}-\chi''^{D_{6h}}_{RL}/2,\\
c^2&=\chi''^{D_{6h}}_{XY}-\chi''^{D_{6h}}_{RL}/2,\\
f^2&=\chi''^{D_{6h}}_{RL}/2.\\
\end{split}
\end{equation}
Therefore, the algebra in Eq.~(\ref{decompossiitonD6hEQ}) can be used to decompose the measured Raman signal into three separate irreducible representations ($A_{1g}$, $A_{2g}$, $E_{2g}$) of point group $D_{6h}$~[Table~\ref{decompositionD6h}].  
                                                                                                                                                                           
This decomposition algebra is a characteristic property of a lattice system with trigonal or hexagonal symmetries, where the three-fold rotational symmetry is preserved~\footnote{Whether six-fold rotational symmetry is preserved or not depends on the system}.

\section{Phonon instability calculation}\label{Phonon_instabilities}  

\begin{figure}[!t] 
\begin{center}
\includegraphics[width=\columnwidth]{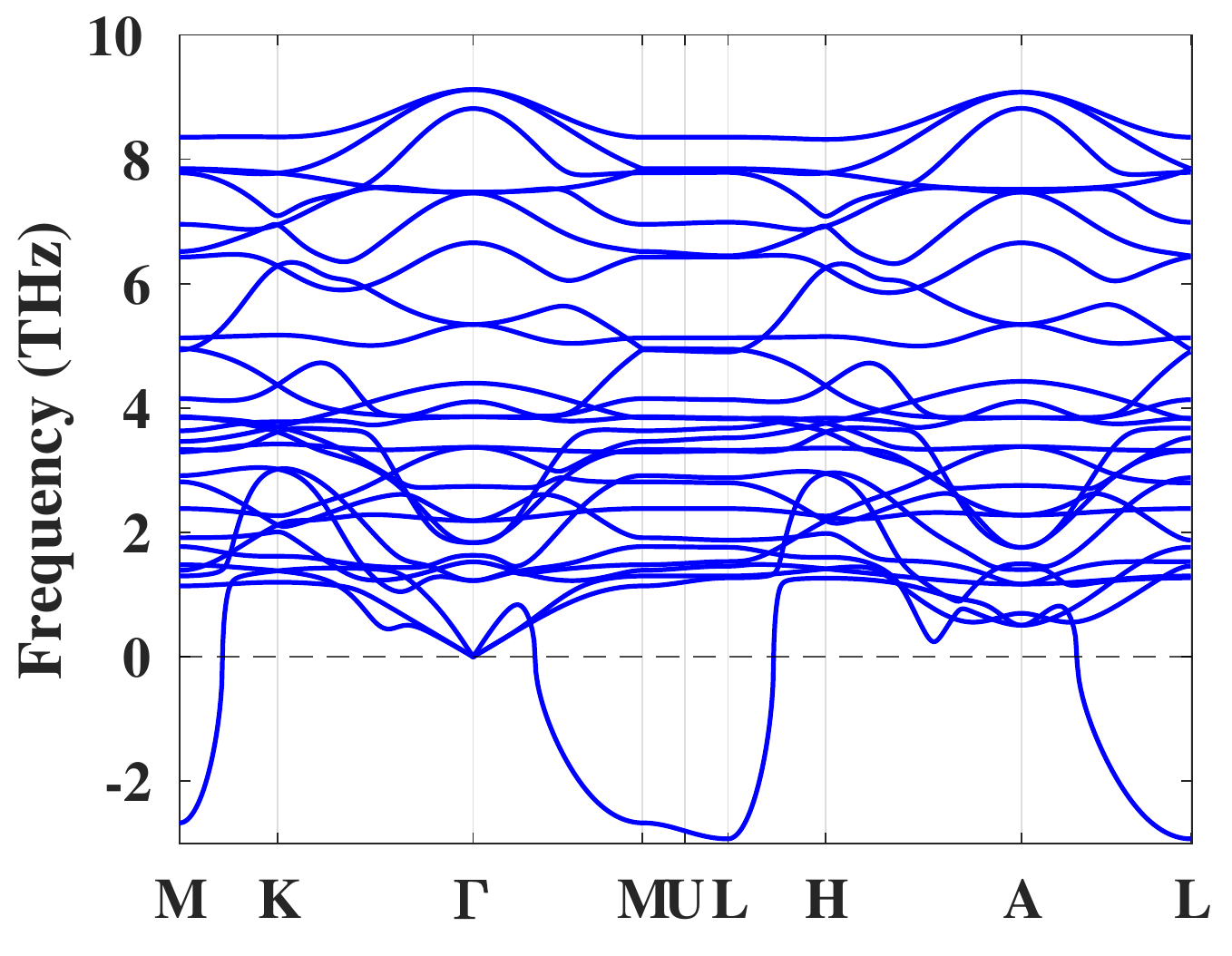}
\end{center}
\caption{\label{CVS_phonon_disperson} 
DFT calculation of phonon dispersion in the whole Brillouin zone for the high-temperature phase of CsV$_3$Sb$_5$.}
\end{figure}   
In Fig.~\ref{CVS_phonon_disperson}, we show DFT calculation of the phonon frequencies in the whole Brillouin zone for the high-temperature phase of CsV$_3$Sb$_5$. 
The lowest phonon branch shows a broad imaginary response along the whole $U$ line [$M$-$U$-$L$ line], as well as the $\Gamma M$ and $AL$ lines. 
Therefore, the DFT calculation does not provide a unique identification of the lattice instability vector.

\section{Group-theoretical analysis}\label{GroupTheory}

In this appendix, we focus on the folded phonon modes in the BZ center which originate from BZ boundary due to the CDW ordering. 

Extending the high-temperature supercell to the $2 \times 2 \times 1$ supercell, we have two types of CDW order driven by V  displacements in the kagome layer: SoD and iSoD structure.
These two superlattices share the same space group of $P6/mmm$ as the high-temperature phase, displaying an in-plane $2 \times 2$ modulation of the high-temperature structure. 
From the group-theoretical considerations~\cite{Bilbao_1}, $\Gamma$-point phonon modes of the $2 \times 2 \times 1$ supercell can be expressed as $\Gamma_{\text{total}}$ = 5$A_{1g}$ + $A_{1u}$ + 3$A_{2g}$ + 9$A_{2u}$ + 4$B_{1g}$ + 5$B_{1u}$ + 2$B_{2g}$ + 7$B_{2u}$ + 8$E_{2u}$+ 8$E_{2g}$ + 14$E_{1u}$ + 6$E_{1g}$. Raman active modes $\Gamma_{\text{Raman}}$= 5$A_{1g}$ + 8$E_{2g}$ + 6$E_{1g}$, IR active modes are $\Gamma_{\text{IR}}$=8$A_{2u}$ + 13$E_{1u}$, the acoustic mode $\Gamma_{\text{acoustic}}$ =$A_{2u}$ + $E_{1u}$ and the silent modes are $\Gamma_{\text{silent}}$ =  3$A_{2g}$  + 4$B_{1g}$ + 2$B_{2g}$ + $A_{1u}$ + 5$B_{1u}$ + 7$B_{2u}$ + 8$E_{2u}$. Note that alkali atoms (K, Rb, Cs) do not involve any Raman-active vibration in the $2 \times 2 \times 1$ SoD and iSoD phase, as well as in the high-temperature phase. 
                       
Extending the $2 \times 2 \times 1$ supercell to the three-dimensional $2 \times 2 \times 4$ supercell, 
we have a structure composed of one layer of the iSoD structures and three consecutive layers of SoD structure, with zero-phase-shift between neighboring layers~[Fig.~\ref{Fig6_displacement_pattern}].
It has the same space group $P6/mmm$ as the high-temperature phase. 
From the group-theoretical considerations, $\Gamma$-point phonon modes of this superstructure can be expressed as $\Gamma_{\text{total}}$ = 26$A_{1g}$ + 6$A_{1u}$ + 10$A_{2g}$ + 30$A_{2u}$ + 20$B_{1g}$ + 16$B_{1u}$ + 12$B_{2g}$ + 24$B_{2u}$ + 31$E_{2u}$ + 33$E_{2g}$ + 45$E_{1u}$ + 35$E_{1g}$, where Raman active modes $\Gamma_{\text{Raman}}$= 26$A_{1g}$ + 33$E_{2g}$ + 35$E_{1g}$, IR active modes are $\Gamma_{\text{IR}}$= 29$A_{2u}$ + 44$E_{1u}$, the acoustic mode $\Gamma_{\text{acoustic}}$ =$A_{2u}$ + $E_{1u}$ and the silent modes are $\Gamma_{\text{silent}}$ = 6$A_{1u}$ + 10$A_{2g}$ + 20$B_{1g}$ + 16$B_{1u}$ + 12$B_{2g}$ + 24$B_{2u}$ + 31$E_{2u}$.      
Note that the alkali-metal atoms (Cs, Rb, K) lattice vibrations become Raman-active in the $2 \times 2 \times 4$ superstructure.

\section{Anharmonic phonon decay model}\label{Anharmonic_decay_model}    
In this appendix, we discuss the anharmonic phonon decay model.
We fit the temperature dependence of the phonon frequency and HWHM by anharmonic phonon decay model~\cite{Klemens_PhysRev148,Cardona_PRB1984}: 
\begin{equation}
\label{eq_omega1}
\omega_1(T)=\omega_{0}-C_1\left[1+ 2n(\Omega(T)/2) \right],\\
\end{equation}
\begin{equation}
\label{eq_gamma1}
\Gamma_1(T)=\gamma_{0}+\gamma_1\left[1+ 2n(\Omega(T)/2)\right],
\end{equation}
\begin{equation}
\label{eq_omega2}
\omega_2(T)=\omega_1(T)-C_2\left[1+ 3n(\Omega(T)/3)+3n^2(\Omega(T)/3)\right],
\end{equation}
\begin{equation}
\label{eq_gamma2}
\Gamma_2(T)=\Gamma_1(T)+\gamma_2\left[1+ 3n(\Omega(T)/3)+3n^2(\Omega(T)/3)\right],
\end{equation}
where $\Omega(T)= \hbar \omega / k_BT$, $n(x)=1/(e^x-1)$ is the Bose-Einstein distribution function. $\omega_1(T)$ and $\Gamma_1(T)$ involves mainly three-phonon decay process where an optical phonon decays into two acoustic modes with equal energy and opposite momentum. 
$\omega_2(T)$ and $\Gamma_2(T)$ involves additional four-phonon decay processes compared with $\omega_1(T)$ and $\Gamma_1(T)$.

%

\end{document}